\documentclass[11.5pt, twocolumn]{article}
\usepackage[a4paper, total={6.5in, 9.5in}]{geometry}
\pdfoutput=1

\usepackage{graphicx}
\usepackage{dcolumn}   
\usepackage{bm}        
\usepackage{amssymb}   
\usepackage{float}
\usepackage{xcolor}
\usepackage{soul}
\usepackage[normalem]{ulem}
\usepackage{authblk}
\usepackage{amsmath}
\usepackage{doi}
\usepackage{url}

\newif\ifcomments
\commentstrue
\commentsfalse

\newcommand{\psin}[0]{\psi_\mathrm{N}}
\newcommand{\grad}[0]{\bm\nabla}
\newcommand{\sep}[0]{\mathrm{sep}}
\newcommand{\ped}[0]{\mathrm{ped}}
\newcommand{\ELM}[0]{\mathrm{ELM}}
\newcommand{\heat}[0]{\mathrm{heat}}

\newcommand{\ie}[0]{i.e., }

\newcommand{\ephi}[0]{\bm{\hat{\varphi}}}

\begin{document}

\title{MHD simulations of small ELMs at low triangularity in ASDEX Upgrade}

\author[1]{\large A.~Cathey\thanks{andres.cathey@ipp.mpg.de}} 
\author[1]{\large M.~Hoelzl}
\author[2]{\large G.~Harrer}
\author[1]{\large M.G.~Dunne}
\author[3,4]{\large G.T.A.~Huijsmans} 
\author[1]{\large K.~Lackner}
\author[5]{\large S.J.P.~Pamela} 
\author[1]{\large E.~Wolfrum}
\author[1]{\large S.~G\"unter}
\author[6]{\large the JOREK team} 
\author[7]{\large the ASDEX Upgrade Team} 
\author[8]{\large the EUROfusion MST1 Team} 

\affil[1]{\small Max Planck Institute for Plasma Physics, Boltzmannstr.2, 85748 Garching, Germany}
\affil[2]{\small Institute of Applied Physics, TU Wien, 1040 Vienna, Austria}
\affil[3]{\small CEA, IRFM, 13108 Saint-Paul-Lez-Durance, France}
\affil[4]{\small Eindhoven University of Technology, P.O. Box 513, 5600 MB Eindhoven, The Netherlands}
\affil[5]{\small CCFE, Culham Science Centre, Abingdon, Oxon, OX14 3DB, United Kingdom}
\affil[6]{\small see the author list of [M. Hoelzl et al 2021 Nucl. Fusion 61 065001]}
\affil[7]{\small see the author list of H. Meyer et al. 2019 Nucl. Fusion 59 112014}
\affil[8]{\small see the author list of B. Labit et al. 2019 Nucl. Fusion 59  0860020}

\date{}
\maketitle

\begin{abstract}
The development of small- and no-ELM regimes for ITER is a high priority topic due to the risks associated to type-I ELMs. By considering non-linear extended MHD simulations of the ASDEX Upgrade tokamak with the JOREK code, we probe a regime that avoids type-I ELMs completely provided that the separatrix density is high enough. The dynamics of the pedestal in this regime are observed to be qualitatively similar to the so-called quasi-continuous exhaust (QCE) regime in several ways. Repetitive type-I ELMs are substituted by roughly constant levels of outwards transport caused by peeling-ballooning modes (with dominant ballooning characteristics) which are localised in the last 5\% of the confined region (in normalised poloidal flux). The simulated low triangularity plasma transitions to a type-I ELMy H-mode if the separatrix density is sufficiently reduced or if the input heating power is sufficiently increased. The stabilising factors that play a role in the suppression of the small ELMs are also investigated by analysing the simulations, and the importance of including diamagnetic effects in the simulations is highlighted. By considering a scan in the pedestal resistivity and by measuring the poloidal velocity of the modes (and comparing to theoretical estimates for ideal and resistive modes), we identify the underlying instabilities as resistive peeling-ballooning modes. Decreasing the resistivity below experimentally-relevant conditions (i.e., going towards ideal MHD), the peeling-ballooning modes that constrain the pedestal below the type-I ELM stability boundary display sharply decreasing growth rates.
\end{abstract}

\section{Introduction}

The thermonuclear experimental fusion reactor ITER is foreseen to operate in high-confinement mode (H-mode), which is characterised by the quasi-periodic excitation of type-I ELMs (edge localised modes)~\cite{Loarte_2014}. In H-mode there are reduced levels of turbulent transport in the edge of the confined region, thus forming a narrow transport barrier that creates a `pedestal' in the density and temperature profiles. Type-I ELMs are macroscopic magnetohydrodynamic (MHD) instabilities that are destabilised by high pressure gradient and/or high toroidal current density (and current density gradient). In standard ELMy H-mode, the pedestal rises (thus increasing $\nabla p$ which simultaneously increases $j_\mathrm{tor}$ by the formation of the bootstrap current) until type-I ELMs are excited. The appearance of type-I ELMs causes the pedestal to crash in a timescale of ${10^{2}\sim10^{3}~\mathrm{\mu s}}$, and it is followed by a quiet inter-ELM phase that lasts ${10^1
\sim10^3~\mathrm{ms}}$. The large transient heat loads associated to these ELMs must be avoided in future tokamaks in order to achieve an acceptable divertor lifetime~\cite{Eich2017A,Gunn_2017}. 

Naturally ELM-free (e.g., QH-mode, EDA H-mode) and ELM-mitigated (e.g., RMP mitigation, pellet pacing, grassy ELMs, small ELMs) operational conditions have been successfully achieved using several methods in different existing tokamaks. However, naturally ELM-free regimes and ELM-mitigated scenarios are only achievable in reduced parameter regimes that differ between different tokamaks and, therefore, it is uncertain which methods and regimes will be accessible in future tokamaks~\cite{Viezzer_2018}. Extrapolating such regimes to ITER becomes increasingly uncertain because ITER parameters cannot be simultaneously accessed with existing tokamaks (e.g., density or collisionality, not both at the same time)~\cite{Oyama_2006}. 

Certain small ELM regimes are completely free of type-I ELMs, maintain other desirable features, and could, if accessible, signify an attractive option for ITER. One such regime is the quasi-continuous exhaust (QCE) regime, which is routinely operated in ASDEX Upgrade (AUG) and TCV and completely avoids type-I ELMs while maintaining good confinement properties~\cite{Harrer_2018,Labit_2019}. In order to access the QCE regime, it is necessary to operate at high separatrix density (${n_{e,\sep}\gtrsim 0.3\times n_\mathrm{GW}}$), with high triangularity and close to double null~\cite{Labit_2019}. The physical mechanism that constrains the pedestal beneath the type-I ELM stability boundary remains unclear, but it is thought to be high-$n$ ballooning modes that are located near the separatrix which cause sufficient outward transport. These modes give rise to a quasi-continuous exhaust of heat and particles which impinge onto the divertor. It has been found that the power fall-off length is larger during the QCE regime than expected from the empirical Eich scaling~\cite{Faitsch_2021}. Other small ELM regimes that can be free of type-I ELMs include grassy ELMs (observed in {JT-60U}~\cite{Kamada_2000}, JET~\cite{Saibene_2005}, EAST and it is foreseen as a potential operational regime for CFETR~\cite{XuGS_2019}), type-III ELMs (not reactor-relevant because they cause confinement degradation~\cite{Sartori_2004}), and a low density small ELM regime in JET~\cite{Garcia_2021}. 

Naturally ELM-free operation includes regimes like QH-mode (in DIII-D~\cite{Burrell_2001}, AUG and JET with carbon wall~\cite{Suttrop_2003,Solano_2010}, and JT-60U~\cite{Sakamoto_2004}), I-mode (features a pedestal in the temperature but not in the density profile~\cite{Viezzer_2018}), and EDA H-mode (found in Alcator C-mod~\cite{Hubbard_2001}). The latter, i.e., the enhanced D-alpha H-mode, was recently achieved in AUG~\cite{Gil_2020}. First experiments of EDA H-mode in AUG were performed at low triangularity with pure electron heating and observed a narrow operational window in terms of the applied heating power. Recently it has been found that the operational window can be extended to higher heating powers by increasing the plasma triangularity; heating above said operational window results in a type-I ELMy H-mode~\cite{Gil_2021eps}. This operational regime always features an edge quasi-coherent mode (in a frequency range between $20$ and ${80~\mathrm{kHz}}$).

The JOREK non-linear extended MHD code~\cite{Hoelzl2021,Huysmans2007} has been extensively used to simulate macroscopic edge instabilities in tokamaks plasmas. In particular for AUG, it has been used to produce realistic simulations of type-I ELMs~\cite{Cathey2020,Hoelzl2018}, RMP-ELM mitigation and suppression~\cite{Orain2019}, and pellet-triggered ELMs~\cite{Cathey_2021,Futatani_2021}. The present article details JOREK simulations of small ELMs at low triangularity in AUG and discusses their relation to the small ELMs that underlie the QCE regime and to the QCM of EDA H-mode. The simulations presented here follow from the approach for modelling the pedestal build-up described in Ref.~\cite{Cathey2020}. In simulations at sufficiently high separatrix density (${n_{e,\sep}\approx0.4\times n_{GW}}$), small ELMs appear beneath the type-I ELM stability boundary and feature medium-$n$ resistive peeling-ballooning modes near the separatrix which cause quasi-continuous heat exhaust. The present article is structured as follows. A brief description of different types of small ELMs in AUG and some features of the EDA H-mode are presented in section~\ref{sec:AUGsmallELMs}. The JOREK model used for the present simulations together with the simulation set-up details and the axisymmetric pedestal build-up are presented in section~\ref{sec:smallELM_setup}. In section~\ref{sec:smallELM_n!0} the results of the non-axisymmetric simulations are presented and a detailed analysis is provided. Two different paths to leave the small ELM regime and reach a type-I ELMy H-mode are presented in section~\ref{sec:smallELMs_typeI}. Finally, conclusions and outlook for future work are discussed in section~\ref{sec:smallELM_conclusions}.

\section{Small ELMs and EDA H-mode at ASDEX Upgrade}\label{sec:AUGsmallELMs}

Small/no ELM scenarios feature different transport mechanisms that cause losses below a few percent of the plasma stored energy. Type-III ELMs are observed in AUG as distinct peaks in the $D_\alpha$ signal with a roughly constant frequency. Each event causes an expulsion of ${\lesssim5\%}$ of the plasma stored energy and their repetition frequency decreases with increasing heating power (${f_\ELM\propto1/P_\heat}$); their repetition frequency is typically larger than type-I ELMs, ${f_\mathrm{type-III}\sim10^3~\mathrm{Hz}}$. They can be obtained either at low pedestal density close to the L-H power threshold or at higher heating power by increasing the pedestal density~\cite{Sartori_2004,Suttrop_2000}. Type-III ELMs are thought to be resistive instabilities, and they are associated with poor confinement properties. JET-like simulations of repetitive edge instabilities that featured an inverse dependency between repetition frequency and heating power have been achieved with JOREK~\cite{Orain2015}. 

The term ``small ELMs'' has been used in AUG as a broad category that includes small amplitude ELMs but excludes type-III ELMs. In particular, this considers type-II ELMs and the ELMs that underlie the QCE regime. The QCE regime is posited to be an attractive scenario for ITER because it completely avoids type-I ELMs while maintaining good confinement properties. Another favourable feature of this regime is that it deposits the expelled energy in a quasi-continuous manner and in a broader area than observed in the inter type-I ELM phase~\cite{Faitsch_2021}. Small ELMs act as a transport mechanism that expels heat and particles such that the pedestal cannot build-up to a point where type-I ELMs are excited; it is presently hypothesised that small ELMs are ballooning modes and/or high-$n$ peeling-ballooning modes that are located at, or very near, the magnetic separatrix. The important ingredients for maintaining the QCE regime are high separatrix density (${n_{e,\sep}/n_\mathrm{GW}\gtrsim0.3}$) and closeness to double null (together with high triangularity)~\cite{Harrer_2018,Labit_2019,Stober_2001}. Because there is little variation of separatrix temperature in a given device (${T_{e,\sep}\approx100~\mathrm{eV}}$ for H-modes in AUG~\cite{Neuhauser_2002}), high $n_{e,\sep}$ translates to high separatrix collisionality (${\nu_e^*\propto n_e/T_e^2}$). In existing tokamaks, high ${\nu^*_{e,\sep}}$ implies also high pedestal collisionality because the temperature cannot increase arbitrarily due to the excitation of ELMs. On the other hand, ITER can reach higher pedestal top temperatures and is expected to operate with high ${\nu^*_{e,\sep}}$ and low ${\nu^*_{e,\mathrm{ped}}}$. Such conditions cannot be achieved simultaneously in existing tokamaks; therefore, fundamental uncertainties exist on whether ITER could operate in the QCE regime~\cite{Harrer_2018}. 

\begin{figure}[!h]
    \centering
    \includegraphics[width=0.48\textwidth]{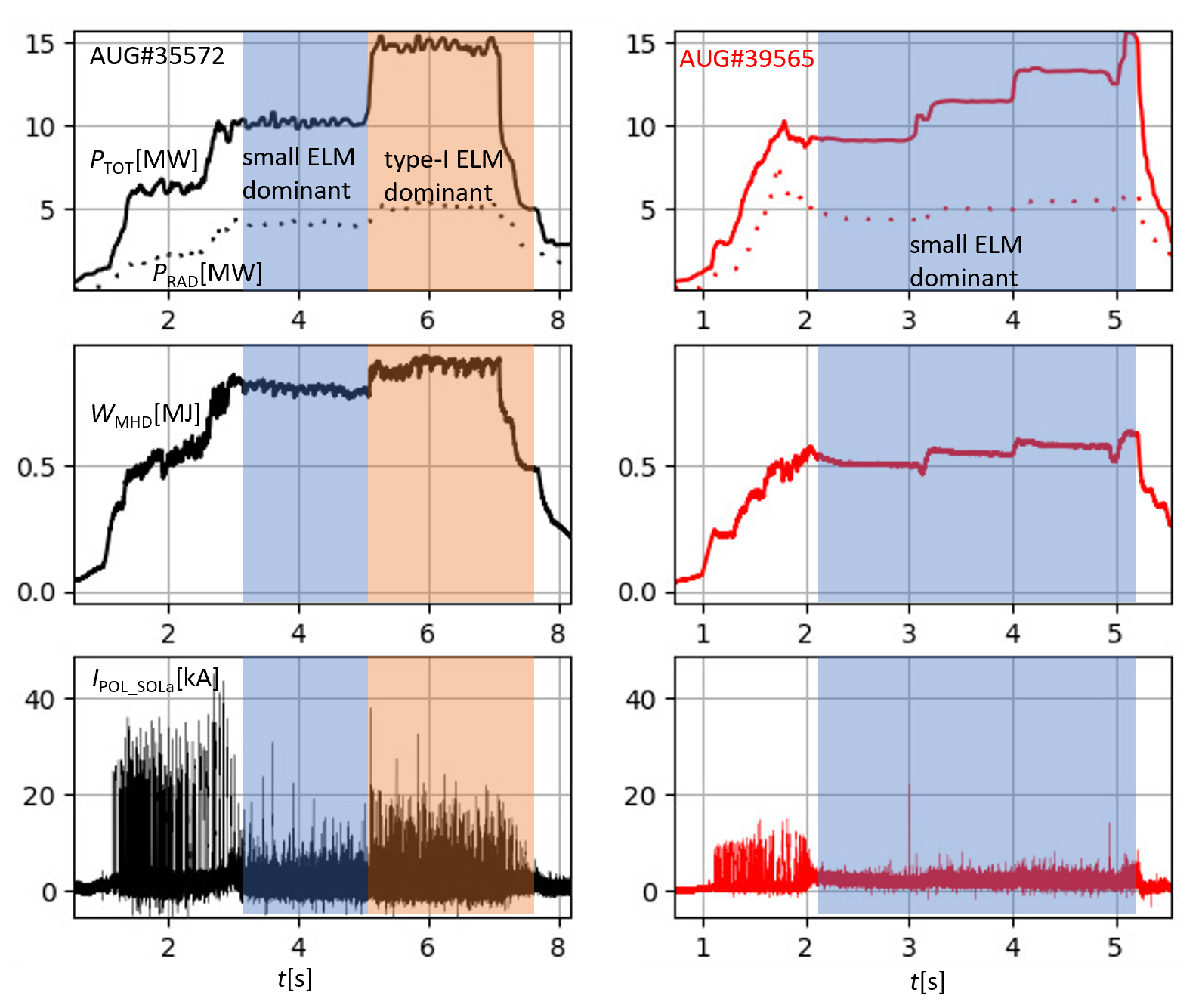}
    \caption{From top to bottom, the heating power (radiated power in dotted lines), plasma stored energy, and ELM monitor of two AUG discharges with small ELMs that feature increasing heating power. A discharge with low edge safety factor (left; \#35572, ${q_{95}=3.73}$) changes from a small ELM-dominant regime to a type-I ELM-dominant regime upon increasing the heating power. The discharge with high ${q_{95}}$ (right; \#39565, ${q_{95}=5.76}$), on the other hand, remains in a small ELM-dominant regime throughout the heating power steps.}
    \label{fig:augcompare}
\end{figure}

Depending on the triangularity and the edge safety factor, increasing heating power can either lead to a sustained small ELM regime (with ``standard'' QCE parameters: high triangularity, high $q_{95}$, and close to double null) or to a transition from small ELMs to type-I ELMs (with ``standard'' QCE parameters, but lower $q_{95}$), as shown in fig.~\ref{fig:augcompare}. For instance, discharge \#35572 (${-2.0~\mathrm{T}}$, ${1.0~\mathrm{MA}}$, ${\delta=0.397}$, low safety factor ${q_{95}=3.73}$, ${P_\heat=10~\mathrm{MW}}$) heated with ICRH and NBI (${P_\mathrm{rad}\approx4.1~\mathrm{MW}}$) shows small ELMs, and transitions to a type-I ELMy H-mode upon increasing the heating power to ${P_\heat=14.6~\mathrm{MW}}$ (${P_\mathrm{rad}\approx5.2~\mathrm{MW}}$). Conversely, at higher $q_{95}$, discharge \#39565 (${-2.5~\mathrm{T}}$, ${0.8~\mathrm{MA}}$, ${\delta=0.401}$, ${q_{95}=5.76}$) heated with ECRH and NBI retains the pure small ELM behaviour through a stepped increase of the heating power up to ${P_\heat=13~\mathrm{MW}}$ (${P_\mathrm{rad}\approx5.5~\mathrm{MW}}$).

Even with a single null configuration and low triangularity, small ELMs can be achieved at sufficiently high separatrix density. Nevertheless, such small ELMs are associated to degrading confinement and even to an H-mode density limit (${n_e\lesssim n_\mathrm{GW}}$). In such cases as the separatrix density is increased (by increasing the gas puff rate), filamentary transport also increases~\cite{Bernert_2014,Griener_2020} and can lead to a flattening of the pressure gradient. This, in turn, causes a reduction of the edge radial electric field\footnote{The radial electric field well in the pedestal region associated to the edge transport barrier roughly follows ${E_{r,\mathrm{neo}}\approx(e n_i)^{-1}\nabla p_i}$~\cite{Viezzer2014}} which is what causes the back transition to L-mode, i.e., the H-mode density limit (HDL)~\cite{Bernert_2014}.  Recently, a correlation has been observed between ballooning stability at the separatrix and the onset of the HDL in AUG~\cite{Eich_2018} and in JET-ILW~\cite{Sun_2021}. In this context, the deterioration of the H-mode confinement and, ultimately, the breakdown of the H-mode are explained by an excess of cross-field transport caused by small ELMs located near the separatrix. 

The enhanced $D_\alpha$ H-mode is a no-ELM operational scenario with high density which is potentially attractive for ITER. Type-I ELMs are not destabilised during EDA H-mode operation because the pedestal is not able to build-up sufficiently. The transport mechanism that allows the pedestal to remain below the type-I ELM stability boundary is thought to be an electromagnetic mode dubbed quasi-coherent mode (QCM)~\cite{Mossessian_2002}. At low triangularity, the EDA H-mode lives in a very narrow operational space in terms of the applied heating power; however, in AUG this operational space has been observed to expand when increasing the plasma triangularity~\cite{Gil_2021eps}. Nonetheless, a stationary EDA H-mode inevitably transitions to a type-I ELMy H-mode upon a sufficient increase of the heating power. In AUG, the QCM moves in the electron diamagnetic direction and has fluctuation frequency in the range ${f\approx20-80~\mathrm{kHz}}$~\cite{Gil_2020}. Similarly, type-II ELMs are accompanied by broadband fluctuations (in magnetic pick-up coils and ECE signals) with frequencies in the range ${f\approx30-50~\mathrm{kHz}}$~\cite{wolfrum2011characterization}.

\section{JOREK simulation set-up and axisymmetric build-up}\label{sec:smallELM_setup}

The simulations presented in this paper were produced with the reduced MHD version of the 3D non-linear extended MHD code JOREK~\cite{Hoelzl2021,Huysmans2007}. Said model simplifies the visco-resistive MHD equations with two considerations. First, the toroidal magnetic field is constrained to be time-independent (${\bm B_\varphi = \frac{F_0}{R}\ephi}$, where $F_0$ is a constant, $R$ is the major radius, and $\ephi$ is the toroidal coordinate). Second, the poloidal velocity is considered to be comprised of the $\mathrm{ExB}$ velocity, which is further considered to lie in the poloidal plane, i.e., ${\bm v_\mathrm{pol} = \bm v_\mathrm{ExB}}$; this assumption allows a potential formulation for the poloidal velocity through the electrostatic potential ${\bm E=-\bm\nabla \Phi}$. It is then possible to include diamagnetic effects (as an extension to the reduced MHD model) by considering the poloidal velocity to be
\begin{align}
    \bm v_\mathrm{pol} &= \bm v_\mathrm{ExB} + \bm v_{*,i} \nonumber \\
    &= -\frac{R}{F_0}\bm\nabla\Phi\times\ephi - \frac{m_i R}{e F_0 \rho}\bm\nabla p_i\times\ephi, \nonumber
\end{align}
where ${m_i}$ and ${p_i}$ are the ion mass and pressure respectively, $e$ is the fundamental electric charge and $\rho$ is the mass density. The resulting extended MHD model is a closed system of five equations for the poloidal magnetic flux ($\psi$), the electrostatic potential ($\Phi$), the parallel velocity ($v_\parallel$), the mass density ($\rho$), and the single fluid temperature ($T$). The inclusion of diamagnetic effects allows the JOREK simulations to recover realistic radial electric fields in the pedestal region, i.e., ${E_r \propto n_i^{-1}\nabla p_i}$~\cite{Viezzer2014}, which play a fundamental role in the stability of PB modes (particularly those with high toroidal mode numbers)~\cite{Rogers_1999}. The stability of PB modes is also partly determined by the edge current density (and its gradient), which is comprised of an Ohmic contribution and a bootstrap current contribution. The former is included in JOREK by assigning a current density source determined by the initial current density profile (which is also comprised by Ohmic and bootstrap current contributions, ${j_0 = j_{0,\Omega}+j_{0,\mathrm{bs}}}$). The time-evolving contribution from the bootstrap current density is considered in JOREK by making use of the Sauter analytical expression~\cite{Sauter1999,Sauter2002}. The total current density source then corresponds to ${j_\mathrm{source}(t) = j_0 + [j_\mathrm{bs}(t) -j_{0,\mathrm{bs}} ] = j_{0,\Omega} + j_\mathrm{bs}(t)}$, and it is included in the induction equation:
\begin{align}
    \frac{1}{R}\partial_t\psi &= - \eta (j-j_\mathrm{source}) - \frac{1}{F_0} [\Phi,\psi] -\frac{1}{R}\partial_\varphi\Phi \nonumber \\
    &\qquad\qquad\,\,\,\,\, + \frac{m_i}{e F_0 \rho}\left( [p_e,\psi] + \frac{F_0}{R} \partial_\varphi p_e \right), \nonumber
\end{align}
where ${[A,B]=(\partial_R A)(\partial_Z B) - (\partial_Z A)(\partial_R B)}$ is the Poisson bracket, and $\eta$ is the resistivity. The extension to include the bootstrap current density as a source term in JOREK was introduced in Ref.~\cite{Pamela2017}, and it has been used in recent simulations of ELMs~\cite{Cathey2020,Cathey_2021,Futatani_2021}. All simulations presented in this paper consider diamagnetic effects and the bootstrap current density source unless specified otherwise. The remainder of the section describes the numerical set-up of the simulations and the axisymmetric build-up of the pedestal profiles.

\subsection{Numerical parameters and simulation set-up}

Initial conditions which are stable to ideal peeling-ballooning modes are considered for this work. These are obtained from a post-ELM equilibrium reconstruction of AUG discharge \#33616 at roughly ${7~\mathrm{s}}$ with the CLISTE code~\cite{McCarthy1999}. The magnetic field at the magnetic axis is $2.5~\mathrm{T}$ and the plasma pressure is ${I_p=0.8~\mathrm{MA}}$. It considers a lower single null magnetic configuration with low triangularity ${\delta_\mathrm{av}=0.29}$ and with the ion ${\bm B\times\grad B}$ drift direction pointing towards the active X-point. The reconstructed current density profile is constrained from based on the steepness of the density, temperature, and pressure profiles, i.e., a bootstrap current constraint~\cite{Dunne_2012}. 

\begin{figure}[!h]
    \centering
    \includegraphics[width=0.48\textwidth]{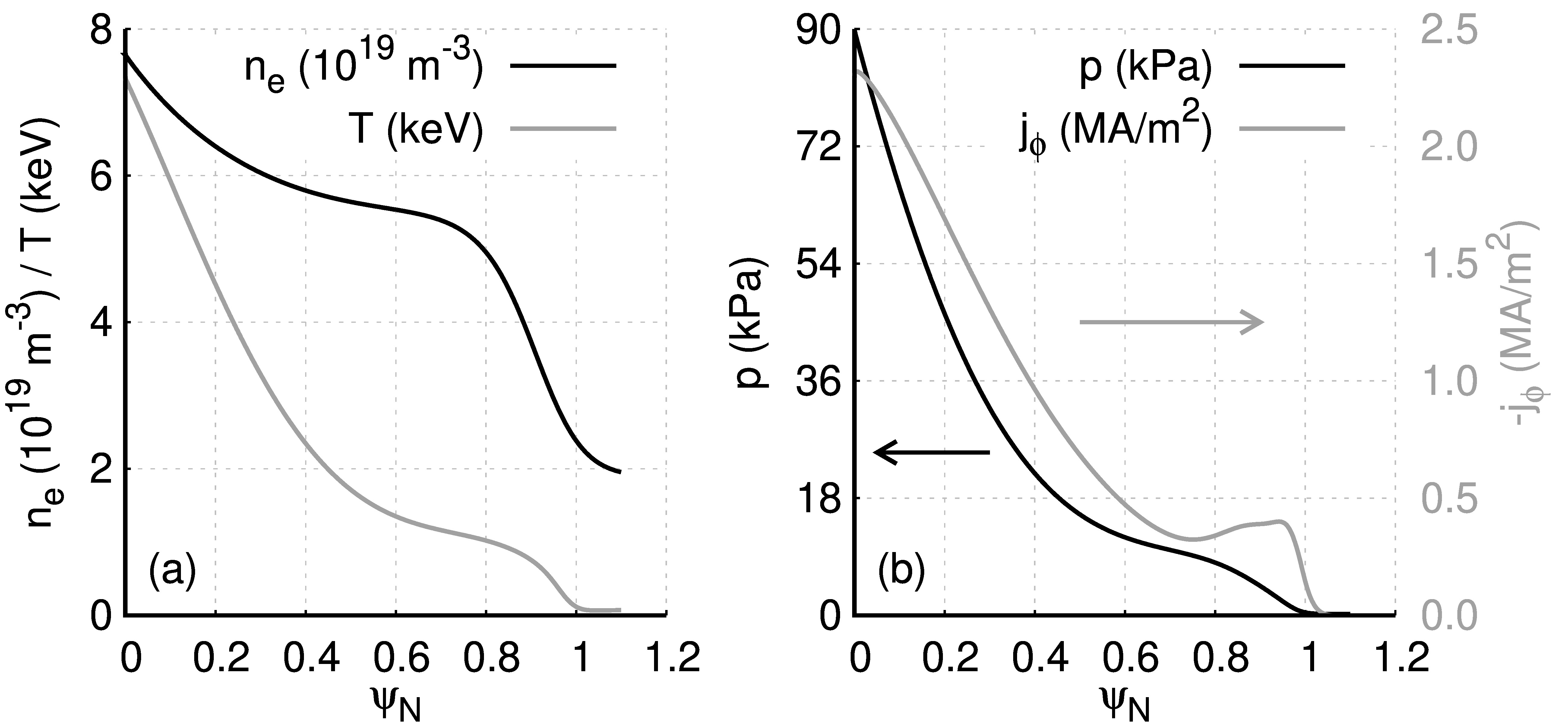}
    \caption{Initial conditions for the electron density and plasma temperature (${T=T_e+T_i}$) (a), and plasma pressure (${p=p_e+p_i}$) and toroidal current density (b) in the outboard midplane.}
    \label{fig:initialConditions}
\end{figure}

The electron density, plasma temperature and pressure, and toroidal current density initial profiles at the outboard midplane are shown in fig.~\ref{fig:initialConditions}. The plasma temperature, $T$, is the sum of the ion and electron temperatures, which are assumed to be equal. The toroidal current density is comprised of an Ohmic contribution together with a Pfirsch-Schl\"uter and a bootstrap current contribution\footnote{The Pfirsch-Schl\"uter current is a force-free current which does not add up in the flux-surface averaged current density, ${\langle j \rangle_{\psin}}$.}. The bootstrap current is only a small contribution in the initial profiles due to the small steepness of the initial pressure profile. An ideal MHD stability analysis with the MISHKA code indicates that these post-ELM pedestal profiles are stable (to ideal PB modes) and would first need to steepen in order to excite a type-I ELM.

We use a flux-surface aligned grid that considers the confined region, the scrape-off layer, and the private flux region. The grid is comprised of 138 points in the radial direction (120 points in the confined region and 18 in the scrape-off layer) and 354 in the poloidal direction. A convergence scan in the grid resolution has been performed; we observe that the linear growth rates of instabilities with ${n\leq20}$ do not change by changing the radial and poloidal resolution (higher mode numbers were not probed because the dominant modes are ${n\leq12}$ due to diamagnetic stabilisation). For the axisymmetric build-up, only one poloidal plane is considered, and for the non-axisymmetric simulations 32 poloidal planes are used in order to simulate the toroidal mode numbers ${n=0,2,4,\dots,12}$ through a Fourier decomposition (and 64 planes for ${n=0,2,4,\dots,20}$). The central resistivity that is used for the simulations is ${\eta=6.6\times10^{-8}~\mathrm{\Omega m}}$. This value is larger than the actual resistivity in the centre (${\eta_\mathrm{Spitzer}\approx2.1\times10^{-9}~\mathrm{\Omega m}}$) and increases with decreasing temperature as it follows the Spitzer temperature dependency, ${\eta\propto T^{-3/2}}$. The true experimental pedestal resistivity is larger than the Spitzer value due to neoclassical effects and $Z_\mathrm{eff}$ (which increases from the core to the edge) such that the resistivity used in the simulations agrees with the experimental values within the error bars in the pedestal region. The parallel electron heat diffusion in the pedestal region of AUG may be estimated with the {Spitzer-H\"arm} expression, ${\chi_{\parallel,\mathrm{SH}^e}=3.6\times10^{29} T_{e,[\mathrm{keV}]}^{5/2}/n_{e,[\mathrm{m^{-3}}]}}$~\cite{Hoelzl2010phd}. For a plasma temperature of ${1~\mathrm{keV}}$ and density of ${5\times10^{19}~\mathrm{m^{-3}}}$, \ie ${\psin\approx0.8}$ in the post-ELM equilibrium shown in fig.~\ref{fig:initialConditions}, ${\chi_{\parallel,\mathrm{SH}^e}\approx 1.27\times10^{9}~\mathrm{m^2/s}}$. For the simulations presented in this paper, the parallel heat diffusion is roughly 15 times lower than the {Spitzer-H\"arm} value; as motivated by the heat-flux limit which can account for a reduction of ${\chi_\mathrm{SH}}$ by a factor of ${15-130}$~\cite{Hoelzl2009}. Fast parallel heat transport acts as a stabilising agent for PB modes~\cite{Xu_2011}.

\subsection{Axisymmetric pedestal build-up}

Together with the post-ELM equilibrium reconstruction of AUG discharge \#33616, a pre-ELM reconstruction from the same discharge was considered. An ideal MHD stability analysis (with the code MISHKA) shows that the pre-ELM profiles are unstable to ideal PB modes. Starting from the post-ELM profiles, we impose \textit{ad-hoc} diffusion coefficients (that describe a well in the pedestal region) and sources that drive the pedestal towards the pre-ELM profiles. The diffusion coefficients and sources are constant in time as the simulation progresses. This evidently results in a pedestal build-up at fixed pedestal width, which is a simplification of what is experimentally observed. Time-evolving diffusion coefficients and sources would require including several key physical effects that are beyond the scope of MHD, and which will be investigated in future work (more on this subject in subsection~\ref{ssec:whats-missing}). 

\begin{figure}[!t]
    \centering
    \includegraphics[width=0.48\textwidth]{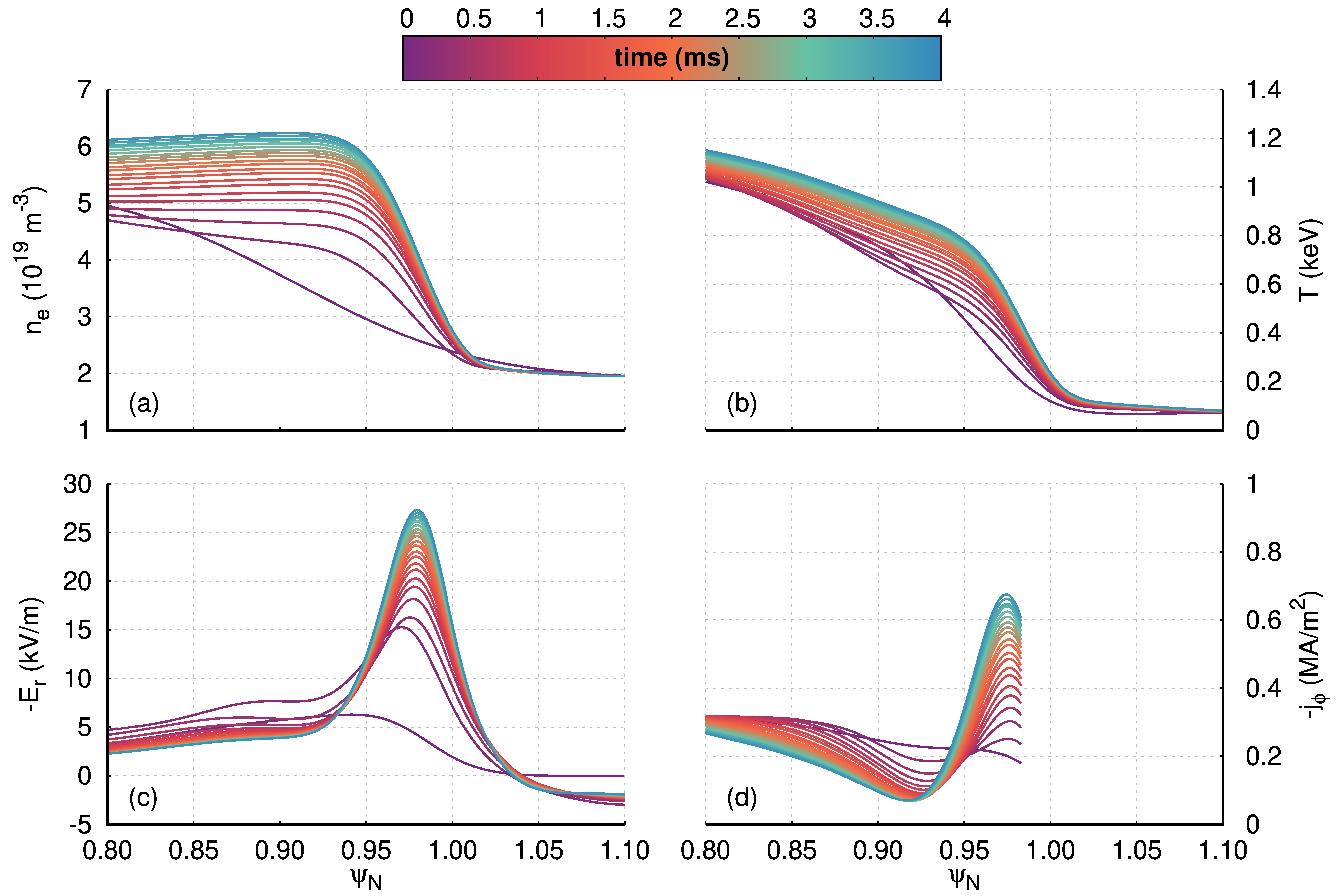}
    \caption{Outboard midplane pedestal profiles of electron density (a), plasma temperature (b), radial electric field (c), and flux-surface averaged toroidal current density (d) during the imposed pedestal build-up. The first millisecond is defined by a strong steepening of the pedestal, and the pedestal top grows progressively at a fixed gradient.}
    \label{fig:smallELM_buildUp}
\end{figure}

As the pedestal $n_e$ and $T$ evolve due to the stationary diffusion and sources, the radial electric field and $j$ become driven by the increasing influence of diamagnetic effects and the bootstrap current density, respectively. Figures~\ref{fig:smallELM_buildUp}(a)-(d) show the time evolution of the pedestal in terms of the electron density, plasma temperature, radial electric field, and the flux-surface averaged toroidal current density. The colours of the profiles change gradually from purple to blue with increasing time as shown in the colour bar on top of the figure. The profiles are plotted every ${0.2~\mathrm{ms}}$ during the first ${4~\mathrm{ms}}$ of an axisymmetric simulation. In the first millisecond, the profiles change shape quickly (as can be most clearly evidenced in the evolution of the density and radial electric field). Section~\ref{sec:smallELM_n!0} will show that non-axisymmetric instabilities driven by the steepening profiles prevent the pedestal from building up towards the profiles shown in fig.~\ref{fig:smallELM_buildUp}. From linear stability analysis performed with the MISHKA code, we know that the pedestal build-up shown in fig.~\ref{fig:smallELM_buildUp} does not cross the ideal peeling-ballooning boundary in the time shown.

\subsection{Limitations of the present approach}
\label{ssec:whats-missing}

The pedestal build-up considered for the simulations presented in this work assumes a constant level of turbulent and neoclassical transport. Experimentally, however, turbulent and neoclassical transport in the pedestal is known to evolve in sub-millisecond and millisecond timescales, respectively. These changes in transport levels would determine how exactly the pedestal top and width evolve. In that sense, to produce a fully realistic pedestal build-up, which would include the pedestal widening, it is necessary to run neoclassical and gyro-kinetic (or even kinetic) simulations to determine the dynamical turbulent and neoclassical transport throughout the simulation time. This would represent not only an extremely costly endeavour from the point of view of computation time, but also would require significant efforts in terms of code development, which lie well beyond the scope of the present work. In the future, neural network based reduced models for the turbulent transport coefficients could potentially be used for incorporating such effects in the MHD simulations (said reduced models, however, do not include transport in the pedestal).

The ion and electron species can be approximated to have the same temperatures and densities in the pedestal region, but in reality ions and electrons behave in distinct ways due to the large difference in their respective masses. Such effects are neglected in the present simulations since we use the single fluid version of the JOREK code. However, a two temperature model has been developed in JOREK and it will be used in the future to understand the effect of such temperature separation in ELM physics and advance the level of realism in our simulations. Another important physical effect that is not considered in the present approach is the penetration of neutral particles onto the confined region and their interaction with the plasma (more generally speaking, a more complete SOL/divertor model is missing). The ensuing ionisation of the neutral particles would determine the amount of particle fuelling that should be considered at any given time during a simulation. These fuelling effects, in turn, directly influence the density (and ultimately temperature) profiles in the pedestal. Ongoing efforts are underway that permit JOREK simulations to consider such effects either by a kinetic treatment~\cite{HuijsmansEPS2019} or a fluid treatment~\cite{Smith2020,Korving_2021eps} of the neutrals.

\section{Non-axisymmetric simulations}\label{sec:smallELM_n!0}

The present section details the simulation results by first describing the linear growth phase of the instabilities together with the early non-linear phase (phase during which ${n\neq0}$ modes interact with each other but not with the ${n=0}$ axisymmetric background), which take place roughly during the first millisecond of simulation time. The poloidal velocities of the linearly unstable modes are measured during this linear growth phase, and the impact of varying the resistivity onto the linear growth rates is probed. These analyses allow us to identify the underlying instabilities as resistive peeling-ballooning modes located near the separatrix. 

\begin{figure}[!h]
    \centering
    \includegraphics[width=0.48\textwidth]{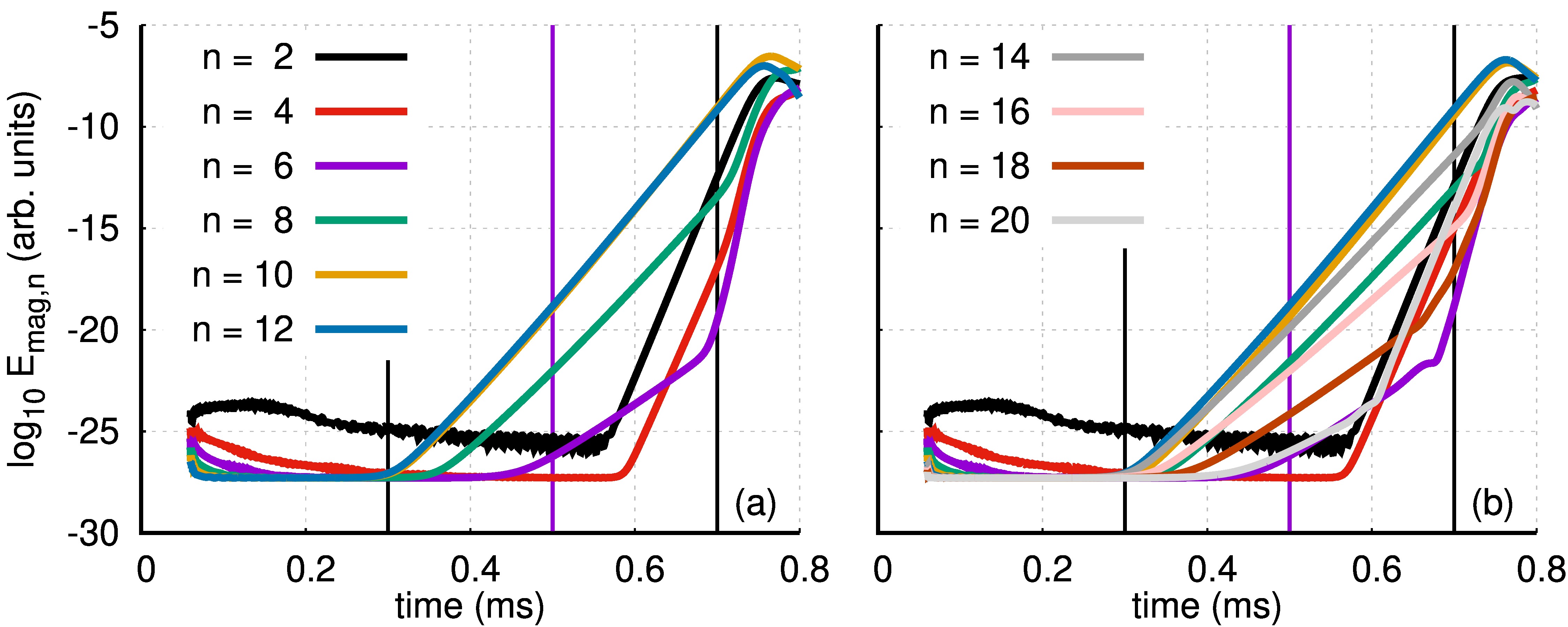}
    \caption{Magnetic energies of the ${n\neq0}$ modes in logarithmic scale in the first ${0.8~\mathrm{ms}}$. (a) shows the simulation with ${n=0,2,4,\dots,12}$ and (b) shows the simulation with ${n=0,2,4,\dots,20}$. This comparison shows that the dominant mode numbers in the linear phase are ${n=10}$ and $12$ in both cases; into the non-linear phase (not shown), the ${n>12}$ modes are sub-dominant.}
    \label{fig:smallELM_linearPhase}
\end{figure}

\begin{figure}[!h]
    \centering
    \includegraphics[width=0.48\textwidth]{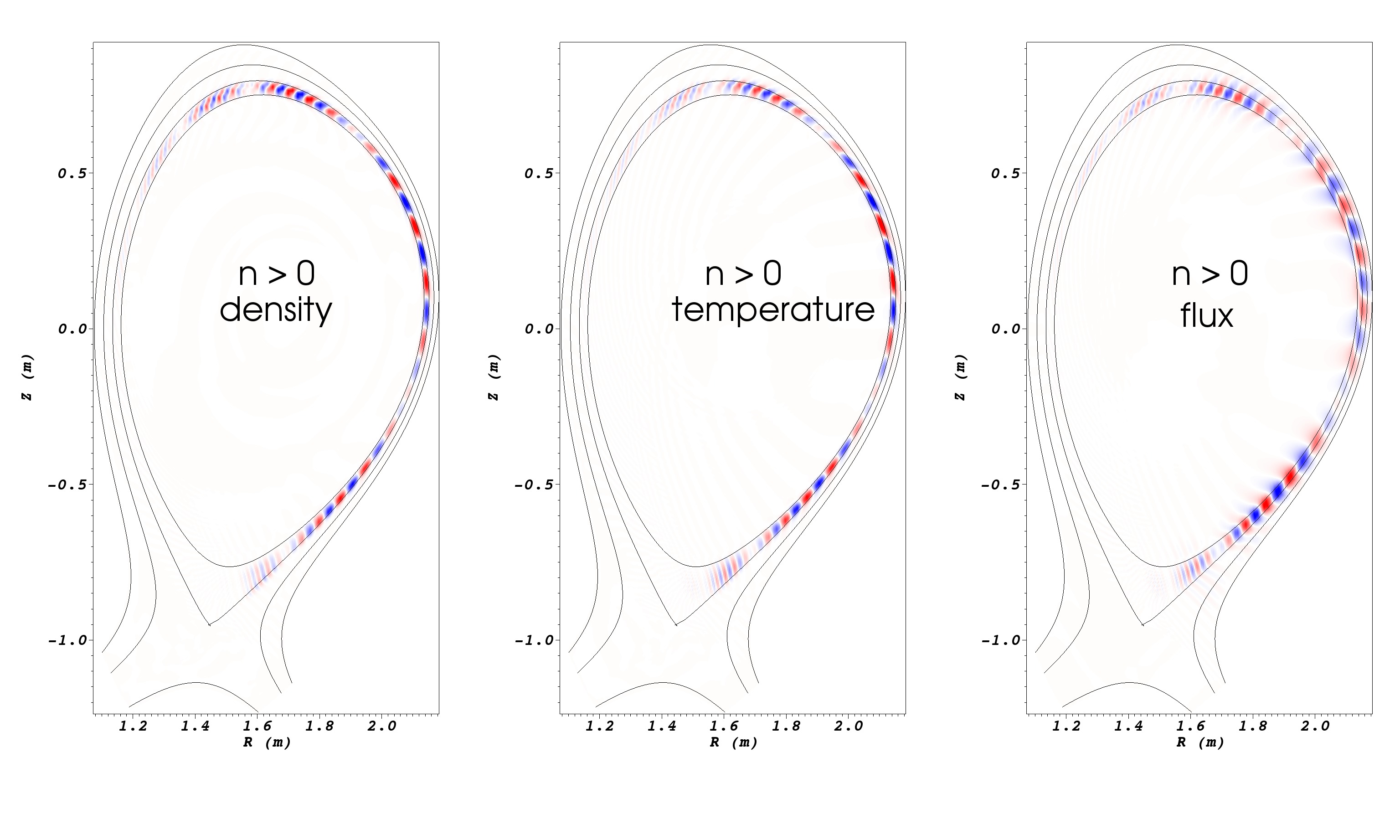}
    \caption{Density, temperature, and poloidal magnetic flux perturbations (${n\neq0}$ components) from the simulation with ${n=0,2,4,\dots,12}$ at ${0.4~\mathrm{ms}}$. The peeling-ballooning structure with dominant ballooning characteristics, \ie more localised to the low field side (LFS), can be observed in both plots. Flux surfaces at ${\psin=0.95,1.00,1.05,}$ and ${1.10}$ are shown with gray lines.}
    \label{fig:smallELM_linearPhasePert}
\end{figure}

During the axisymmetric build-up shown in fig.~\ref{fig:smallELM_buildUp}, the pedestal does not cross the ideal PB boundary (as confirmed by ideal MHD stability analysis with MISHKA). However, non-ideal instabilities can become excited due to the finite resistivity used in JOREK. A simulation which includes the toroidal mode numbers ${n=0,2,4,\dots,12}$ is performed by introducing perturbations with said mode numbers at noise-level amplitudes after ${t\approx0.1~\mathrm{ms}}$ of axisymmetric build-up. After a brief time of stability (${\sim0.2~\mathrm{ms}}$) the steepening pedestal that is not unstable to ideal PB modes begins to excite non-ideal PB instabilities with predominant ballooning features. The magnetic energies of the non-axisymmetric perturbations can be observed in fig.~\ref{fig:smallELM_linearPhase}(a). The time frame between the first vertical black line (${t=0.3~\mathrm{ms}}$) and the vertical purple line (${t=0.5~\mathrm{ms}}$) denotes the linear growth phase where only the linearly unstable modes (${n=6,~8,~10,~\mathrm{and}~12}$) grow; the time frame between the purple line and the second vertical black line (${t=0.7~\mathrm{ms}}$) denotes the early non-linear growth phase where non-linear mode coupling excites linearly stable modes to grow (${n=2}$ and 4)\footnote{An additional simulation with only ${n=2}$ and 4 was ran (not shown) to confirm that ${n=2}$ and 4 are linearly stable and, indeed, no mode growth was observed.}. The structure of the PB modes of fig.~\ref{fig:smallELM_linearPhase}(a) during the linear phase (at ${t=0.4~\mathrm{ms}}$) is shown in fig.~\ref{fig:smallELM_linearPhasePert} with the perturbations of density, temperature, and poloidal magnetic flux. The flux surfaces at ${\psin=0.95,1.00,1.05,1.10}$ are also shown in thin black lines. It is observed that the PB modes are localised very close to the separatrix. The modes rotate in the electron diamagnetic direction (counter-clockwise in fig.~\ref{fig:smallELM_linearPhasePert}); their poloidal velocity will be discussed later.  

The initial growth phase is started by an $n=12$ mode in these simulation and closely followed by the growth of the $n=10$ mode. The growth rates of these modes are very similar, roughly ${\gamma_{n=10,12}\approx5\times10^4\mathrm{/s}}$. A separate simulation including all even toroidal mode numbers until $n=20$ has been produced, and the magnetic energies of the $n\neq0$ modes is shown in fig.~\ref{fig:smallELM_linearPhase}(b). In such a way, we confirm that indeed the fastest growing mode is the $n=12$. The toroidal harmonics with ${n>14}$ grow at a slower rate; they remain sub-dominant well into the non-linear phase (not shown). Quadratic non-linear mode coupling takes place during the early non-linear phase (${t=[0.5,0.7]~\mathrm{ms}}$). The non-linear mode coupling that gives rise to the excitation of linearly stable modes has been described in JOREK simulations from Ref.~\cite{Krebs2013}. The upcoming subsection is devoted to studying the influence of the plasma resistivity onto the growth rates of the non-axisymmetric perturbations.

\subsection{Rigidly scanning the resistivity}

\begin{figure}[!h]
    \centering
    \includegraphics[width=0.48\textwidth]{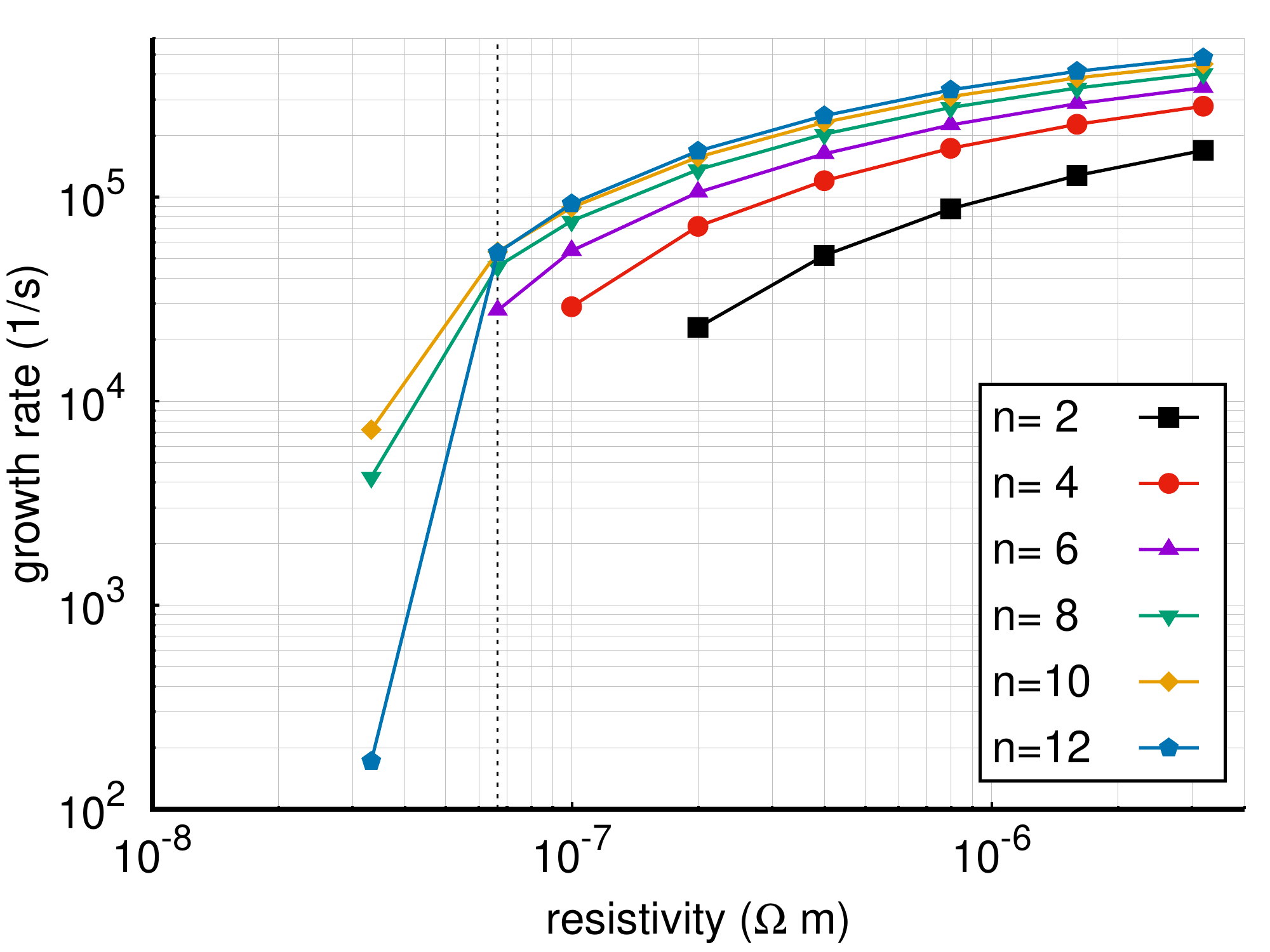}
    \caption{Growth rates of different toroidal mode numbers for the resistivity scan. Increasing the resistivity leads to the destabilisation of resistive PB modes.}
    \label{fig:smallELM_linearPhase-eta}
\end{figure}

In this subsection, we freeze the axisymmetric profiles at ${t=0.5~\mathrm{ms}}$ and change the resistivity to understand its influence onto the stability of the non-ideal PB modes. For this scan we consider several multiplication factors of the nominal resistivity, ${0.5,~1.5,~3.0,~6.0,~12.0,~24.0,~48.0}$. The response of the PB modes to the changes in resistivity display several noteworthy characteristics. The linearly stable modes (${n=2~\mathrm{and}~4}$) at nominal resistivity become linearly unstable by increasing the resistivity (i.e., said modes no longer require non-linear mode coupling to grow). Similarly, decreasing the resistivity by half leads the ${n=6}$ mode to become linearly stable. This is not surprising given the fact that we know (from MISHKA ideal MHD simulations) that ideal PB modes are not unstable for the considered profiles. Varying the resistivity results in a change in the growth rate of the linearly unstable modes, as shown in fig.~\ref{fig:smallELM_linearPhase-eta}. The x-axis represents the resistivity (in ${\mathrm{\Omega~m}}$) and the y-axis corresponds to the growth rate (in $\mathrm{1/s}$); both axes are plotted in logarithmic scales. Different colours and symbols represent different toroidal mode numbers. 

\subsection{Linear growth phase -- mode velocity}

\begin{figure}[!t]
    \centering
    \includegraphics[width=0.48\textwidth]{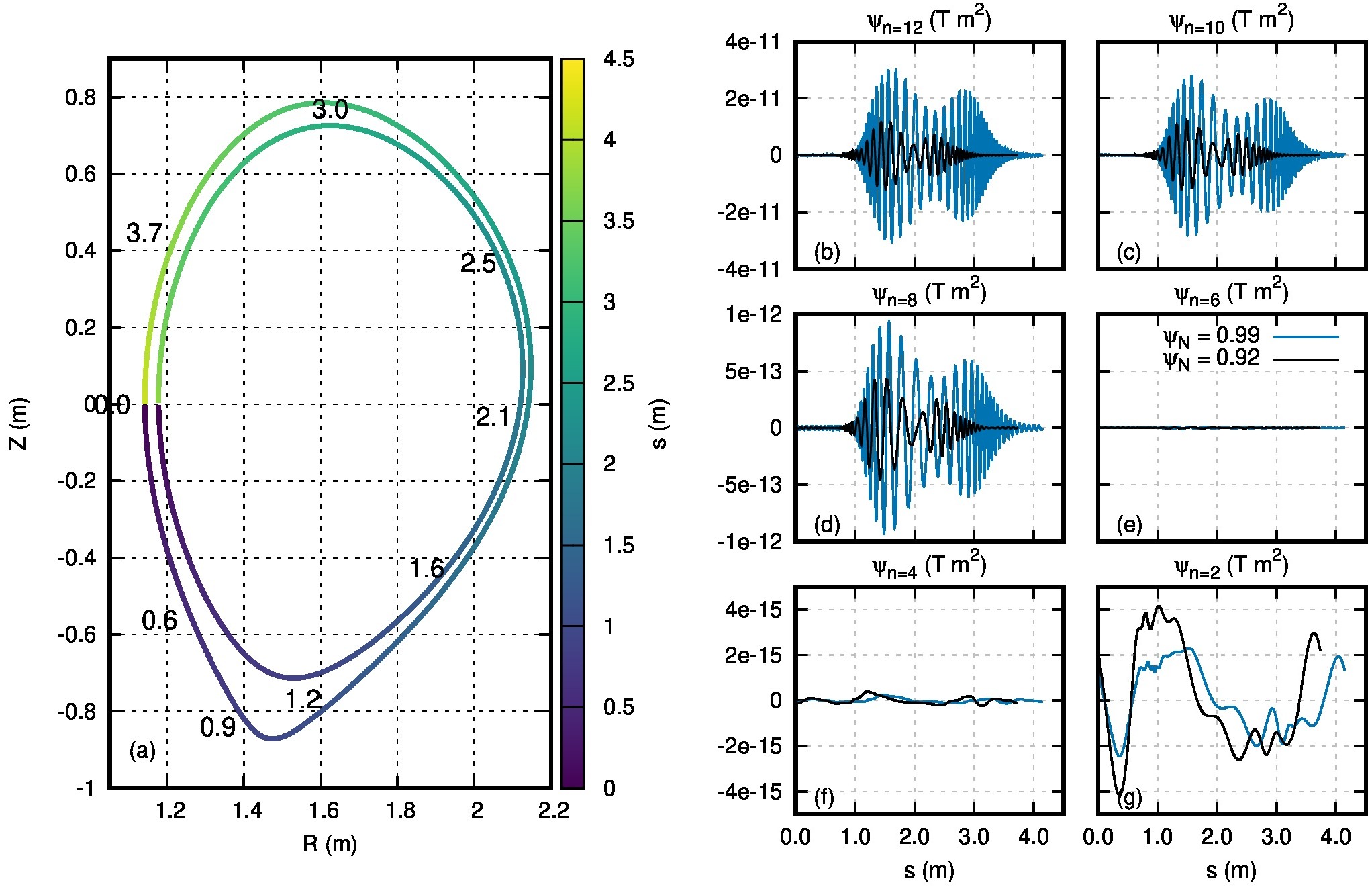}
    \caption{(a) Flux surfaces of ${\psin=0.92}$ and $0.99$ with their corresponding colour coded arc lengths. (b)-(g) The variation of the poloidal magnetic flux perturbations along the arc length for the different toroidal mode numbers. The magnetic flux perturbations for $n=2$ and $4$ do not have any coherent structures as they are linearly stable to the profiles at this point in time (${t=0.5~\mathrm{ms}}$).}
    \label{fig:smallELM_psiPert_s}
\end{figure}

Considering how the peak of the modes in fig.~\ref{fig:smallELM_linearPhasePert} move with time it is possible to determine their poloidal velocity. The modes shown in the figure are the result of perturbations with different toroidal mode numbers; the magnetic flux fluctuation is defined as
\begin{align}
    \Tilde{\psi} = \sum_{n>0}^{n_\mathrm{max}} \psi_n = \psi - \psi_{n=0} , \nonumber
\end{align}
where $n$ is the toroidal mode number. Figure~\ref{fig:smallELM_psiPert_s} shows the ${\psin=0.92}$ and $0.99$ flux surfaces together with the colour coded arc length, $s$, calculated from the inboard midplane (a) and the $\psi_\mathrm{n}$ perturbation for the different toroidal mode numbers (b)-(g) at ${t=0.5~\mathrm{ms}}$. The $n\geq8$ mode amplitudes dominate in the LFS indicating their ballooning nature while the modes with lower $n$ do not have a coherent structure (the $n=6$ already has a coherent structure, but its amplitude is too small to be observed in fig.~\ref{fig:smallELM_psiPert_s}(e)). 

Taking the distance travelled by the peaks of fig.~\ref{fig:smallELM_psiPert_s}(b)-(e) in a small time, it is possible to determine the poloidal velocity of the different modes. 
The poloidal mode velocity of the linearly unstable modes (${n=6,8,10,12}$) at the outer midplane is approximately constant from ${\psin\approx0.92-0.99}$, and corresponds to ${v_{\mathrm{mode},pol}\approx11~\mathrm{km/s}}$.
The poloidal velocity of a mode may be used to attempt to identify the nature of the mode. In particular, from Ref.~\cite{Morales2016}, ideal and resistive ballooning mode rotation velocities in the laboratory frame have been identified as having the following velocities,
\begin{align}
    \mathrm{Resistive:}& \qquad v_{\mathrm{mode},pol} = v_\mathrm{ExB} + v_{\parallel,\mathrm{pol}} , \label{eq:vmode-resistive} \\
    \mathrm{Ideal:}& \qquad v_{\mathrm{mode},pol} = v_\mathrm{ExB} + v_{\parallel,\mathrm{pol}} + v^*_\mathrm{i}/2 , \label{eq:vmode-ideal}
\end{align}
where ${v^*_\mathrm{i}=\nabla p_\mathrm{i}/(e n_e B)}$ is the ion diamagnetic velocity, and ${v_{\parallel,\mathrm{pol}}}$ is the poloidal projection of the parallel velocity. At $0.5~\mathrm{ms}$ of simulation time, eqn.~\eqref{eq:vmode-resistive} results in a poloidal velocity of approximately ${10~\mathrm{km/s}}$ (moving in the electron diamagnetic direction).
On the other hand, eqn.~\eqref{eq:vmode-ideal} results in a poloidal velocity of roughly $4~\mathrm{km/s}$ (in the $e^-$ diamagnetic direction). Comparing these two, it would appear that the mode velocity of the linearly unstable $n=6,8,10,12$ modes is closer to that of resistive ballooning modes than to ideal ballooning modes. Further support for the identification of these modes as resistive modes comes from the fact that their growth rates become larger by increasing the resistivity. Similarly, by reducing the resistivity, the growth rates of the unstable modes decrease, as shown in the previous subsection. Therefore, the unstable high-$n$ modes unstable in the present simulations are characterised as resistive peeling-ballooning modes.

\subsection{Importance of extended MHD}\label{ssec:smallELMs-extendedMHD-importance}

\begin{figure}[!t]
    \centering
    \includegraphics[width=0.48\textwidth]{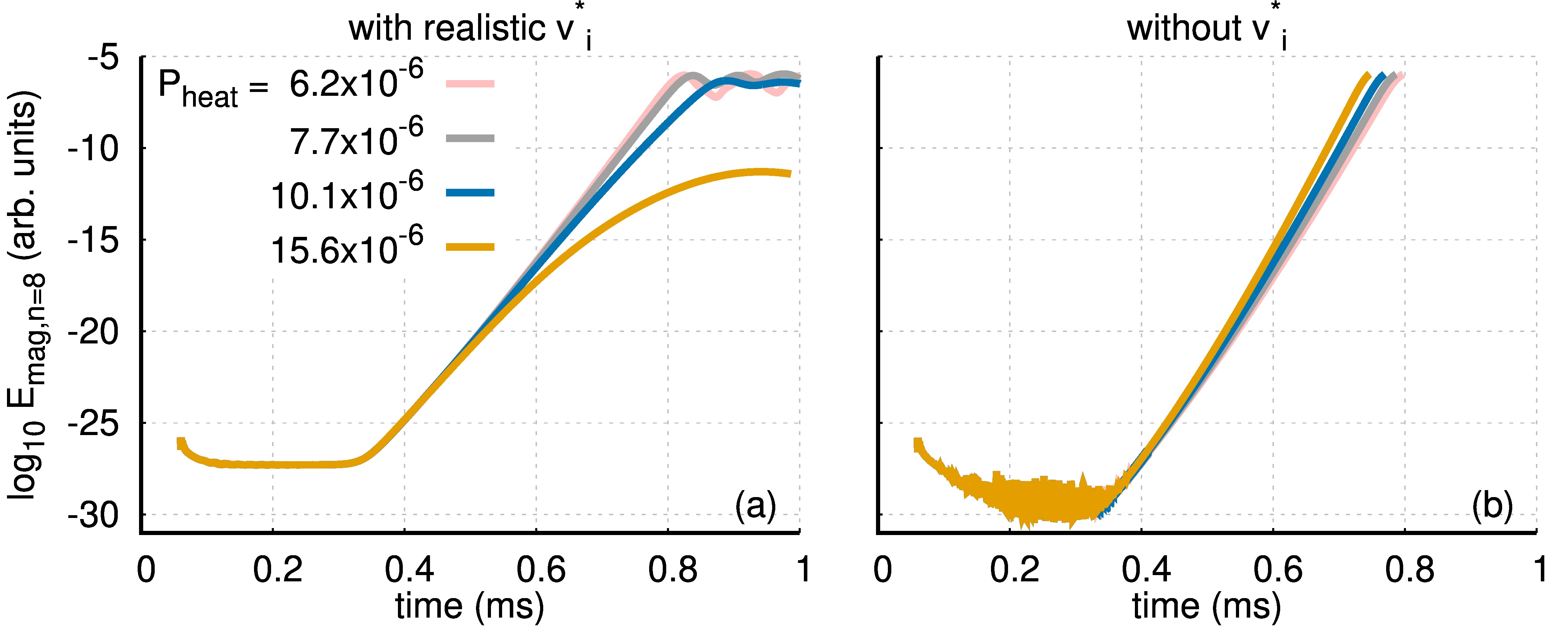}
    \caption{Evolution of ${E_\mathrm{mag,n=8}}$ for simulations with (a) and without (b) the inclusion of ion pressure gradient-driven diamagnetic flows. Four different input heating powers are considered. Therefore, eight single$-n$ simulations are shown. Increasing heating power in simulations that include diamagnetic flows shows an important stabilisation of the $n=8$ PB mode. When neglecting the diamagnetic effects, on the other hand, increasing heating power causes the unstable PB mode to grow even faster due to the steeper pressure profiles.}
    \label{fig:smallELM_diamagEffect}
\end{figure}

Simulations without the diamagnetic effects have been performed in order to understand their influence onto the underlying instabilities described in the previous section. This is done for simulations with only one toroidal harmonic present, ${n=8}$, and for different applied heating powers. It is observed that the simulations that include $v^*_i$ have fundamentally different non-axisymmetric dynamics with respect to the simulations that neglect the diamagnetic effects. Figure~\ref{fig:smallELM_diamagEffect} shows the evolution of the ${n=8}$ magnetic energy in logarithmic scale of (a) simulations with and (b) without diamagnetic effects at four different values of ${P_\mathrm{heat}}$. 

Increasing heating power causes a steepening of the temperature and, therefore, of the pressure at the plasma edge. For the simulations that include diamagnetic effects, the edge ${E_r}$ well at the pedestal becomes deeper with steeper pressure profiles. In said simulations, the PB modes become stabilised by the diamagnetic drift together with the $E_r$ (and its shear)~\cite{Rogers_1999,Huysmans2001,Hastie_2000}. For the simulations without diamagnetic effects the pedestal steepens, but $E_r$ does not change. In fig.~\ref{fig:smallELM_diamagEffect}(a) and (b), the heating power of the different simulations is changed at ${t=0.33~\mathrm{ms}}$ to the values shown in the key. The simulations that consider diamagnetic effects observe ${\gamma_{n=8}}$ to decrease as the heating power is increased. On the other hand, the simulations that neglect diamagnetic effects result in an increase of ${\gamma_{n=8}}$ with increasing heating power. The results presented in this section emphasise the importance of including the diamagnetic flows for simulations of PB modes.

As mentioned at the beginning of this section, the pedestal profiles for the nominal heating power at these stages is stable to ideal peeling-ballooning modes. Higher pedestal pressure and/or edge current densities are required in order to reach a type-I ELM unstable scenario. The access to a type-I ELM unstable scenario appears to be closed without the inclusion of the two-fluid diamagnetic effects. Indeed this result was previously reported in Ref.~\cite{Orain2015} in the context of obtaining repetitive ELM cycle simulations, and was extended as a requirement to simulate type-I ELM cycles in Ref.~\cite{Cathey2020}. The following subsections are devoted to describing the fully non-linear phase (during which the ${n\neq0}$ modes interact with each other and with the ${n=0}$ axisymmetric background) of the simulation with ${n=0,2,4,\dots,12}$ at nominal heating and resistivity in the presence of diamagnetic effects, i.e., fig.~\ref{fig:smallELM_linearPhase}(a).

\subsection{Non-linear phase}\label{ssec:smallELM_nonlinphase}

\begin{figure}[!h]
    \centering
    \includegraphics[width=0.48\textwidth]{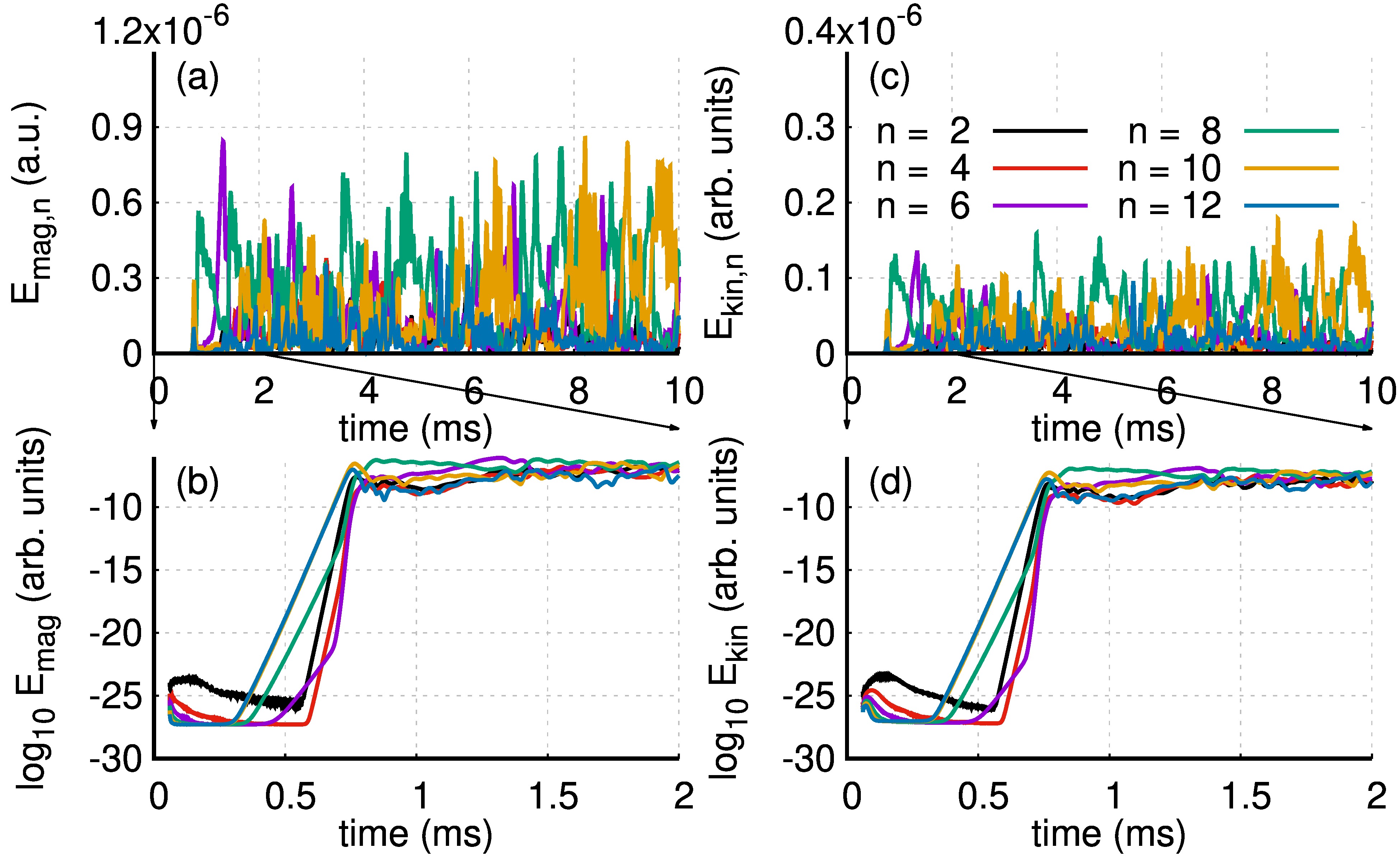}
    \caption{Magnetic and kinetic energies of the non-axisymmetric modes in linear scale for ${10~\mathrm{ms}}$ of simulation time (top) and in logarithmic scale for the first ${2~\mathrm{ms}}$ of simulation time during which the linear growth phase takes place.}
    \label{fig:smallELM_mag_kin_energies1}
\end{figure}

For the simulation that includes ${n=0,2,4,\dots,12}$, \ie figure~\ref{fig:smallELM_linearPhase}(a), the magnetic and kinetic energies of the non-axisymmetric modes are shown in fig.~\ref{fig:smallELM_mag_kin_energies1} in linear scale for ${10~\mathrm{ms}}$ of simulation time (a) and (c) and in logarithmic scale for the first ${2~\mathrm{ms}}$ (b) and (d).

The linear growth phase gives way to the early non-linear growth phase until the amplitude of the perturbations becomes large with respect to the background plasma, at which point the non-linear phase begins. During the latter, a dynamic interplay between the ${n\neq0}$ modes and the background plasma determines the instantaneous profiles observed in the simulations. Due to the persisting PB modes and the lack of a clear cyclical dynamics, the dynamics observed can be characterised as peeling-ballooning turbulence.

\subsection{Filamentary transport}

\begin{figure}[!t]
    \centering
    \includegraphics[width=0.48\textwidth]{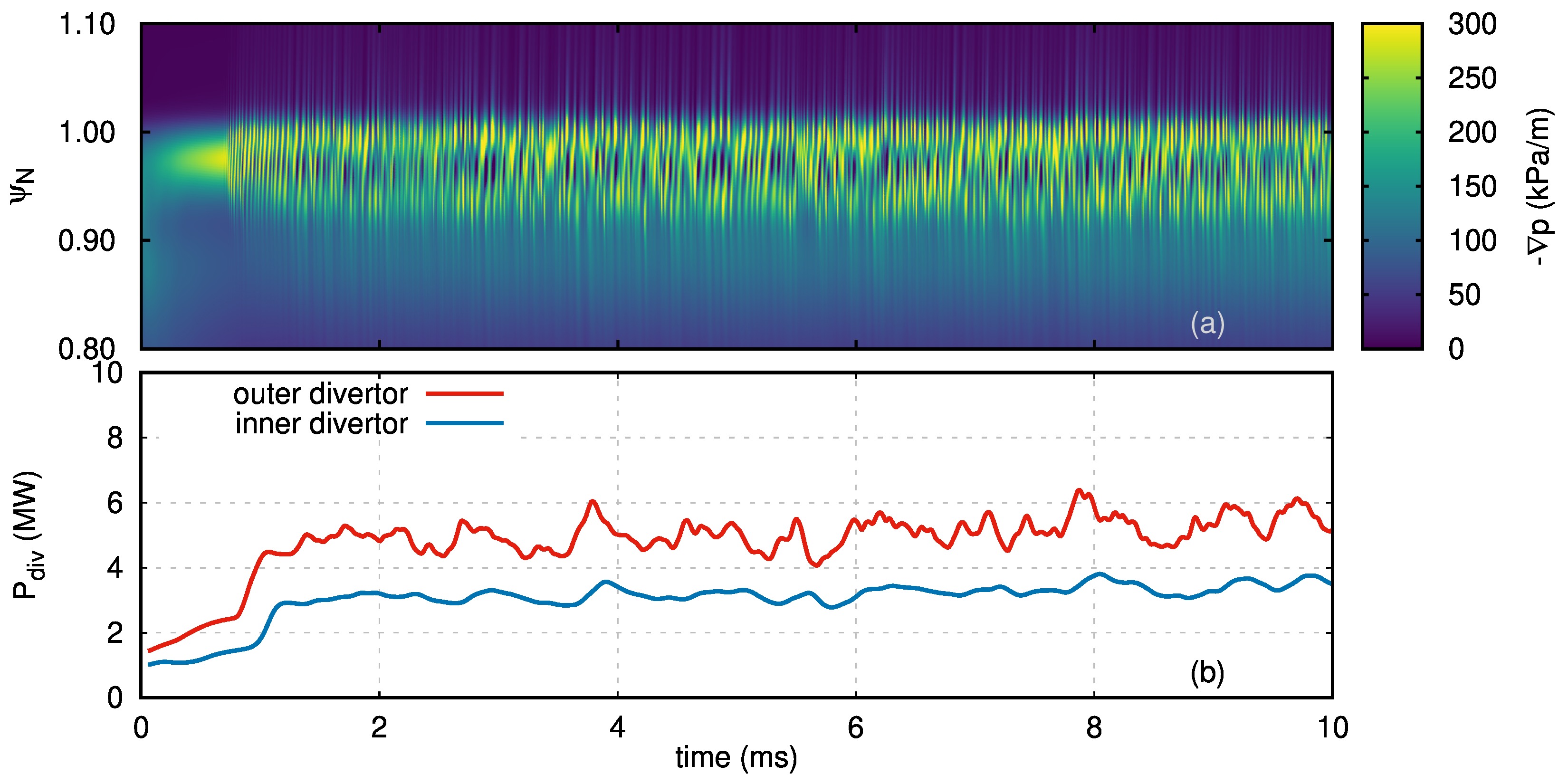}
    \caption{Time evolving outer midplane edge pressure gradient at ${\varphi=0}$ in colour scale (a), and inner/outer divertor incident power (b). The varying pressure profile is caused by quasi-continuous outward transport created by non-ideal peeling-ballooning modes.}
    \label{fig:smallELM_pressure_heat}
\end{figure}

\begin{figure}[!t]
    \centering
    \includegraphics[width=0.48\textwidth]{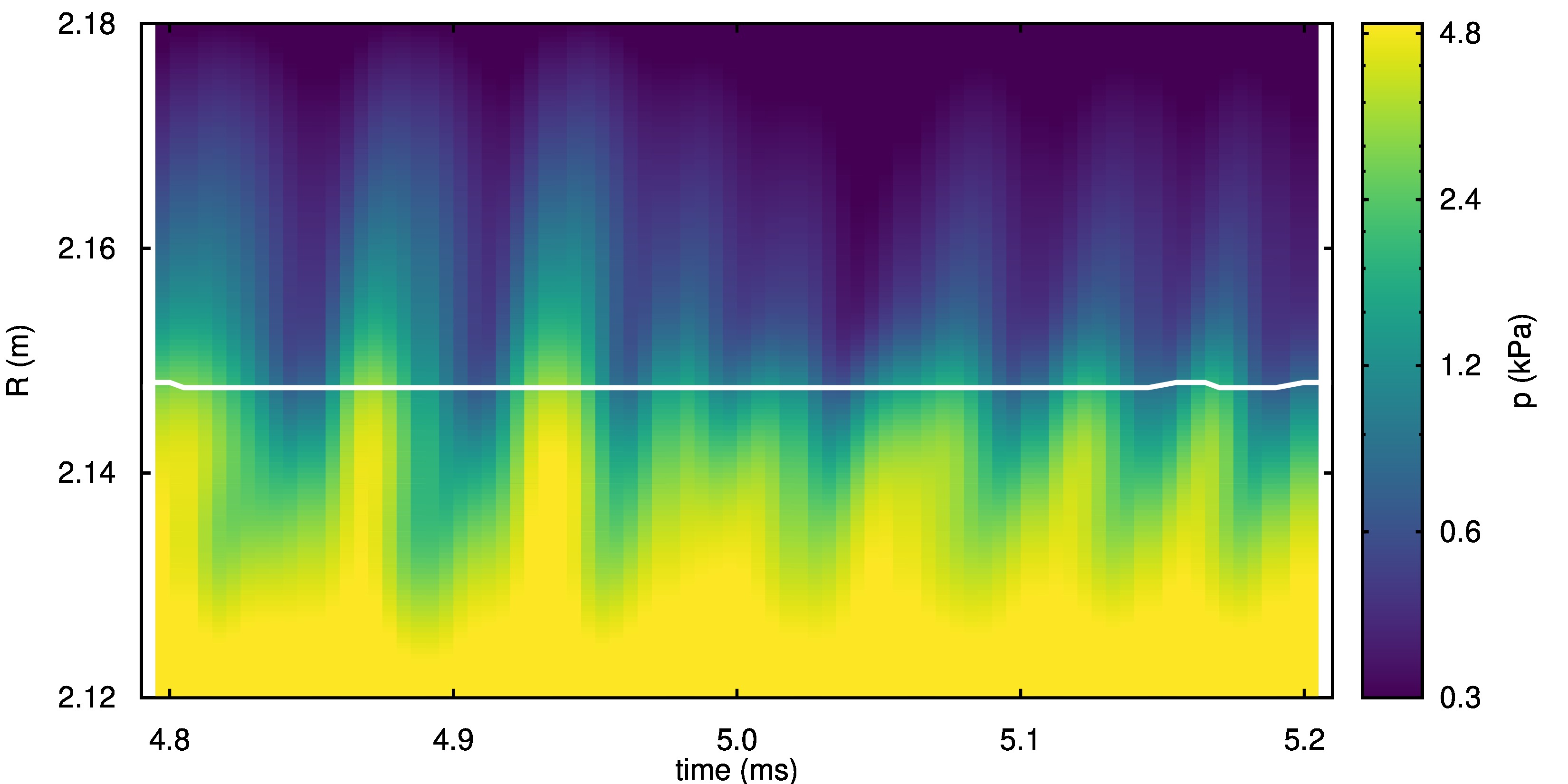}
    \caption{Time evolving outer midplane edge pressure at ${\varphi=0}$ in logarithmic scale. Plasma blobs travelling outwards are aligned to the magnetic field lines.}
    \label{fig:smallELM_pressure_colormap_zoom}
\end{figure}

The non-axisymmetric time evolution of the ${\varphi=0}$ outer midplane pressure gradient and the inner/outer divertor incident power are shown in fig.~\ref{fig:smallELM_pressure_heat}(a) and (b), respectively, for ${10~\mathrm{ms}}$ of simulation time. The incident power is defined as
\begin{align}
    P_\mathrm{div}=\int_0^{2\pi}\int_{\ell_0}^{\ell_\mathrm{max}}q(t,\ell,\varphi) R\,d\ell\,d\varphi , \nonumber
\end{align}
where ${q(t,\ell,\varphi)}$ is the heat flux at a given time in the divertor location $\ell$ at the toroidal angle $\varphi$, and $R$ is the major radius. The colour map indicates that the pressure gradient and, therefore, the pressure profile in the outermost edge of the plasma is rapidly fluctuating. The corresponding fluctuations are governed by the non-axisymmetric modes that regulate the pedestal to fluctuate about a mean value, \ie the PB turbulence. The incident power onto the divertors does not have characteristic spikes, but rather displays a quasi-continuous heat deposition, which is qualitatively similar to the QCE regime or the EDA H-mode in AUG. Figure~\ref{fig:smallELM_pressure_colormap_zoom} shows the ${\varphi=0}$ outer midplane pressure in a colour map with logarithmic scale for a reduced time window of ${0.4~\mathrm{ms}}$, which is chosen between ${4.8}$ and ${5.2~\mathrm{ms}}$, in order to show the dynamics of filamentary structures travelling outwards from slightly inside the separatrix (roughly ${2~\mathrm{cm}}$). The y-axis is the major radius, and the separatrix position is represented with a white line. The plasma blobs that travel outwards result from the resistive PB modes that are aligned to the magnetic fields and are moving in the electron diamagnetic direction.


\begin{figure}[!t]
    \centering
    \includegraphics[width=0.48\textwidth]{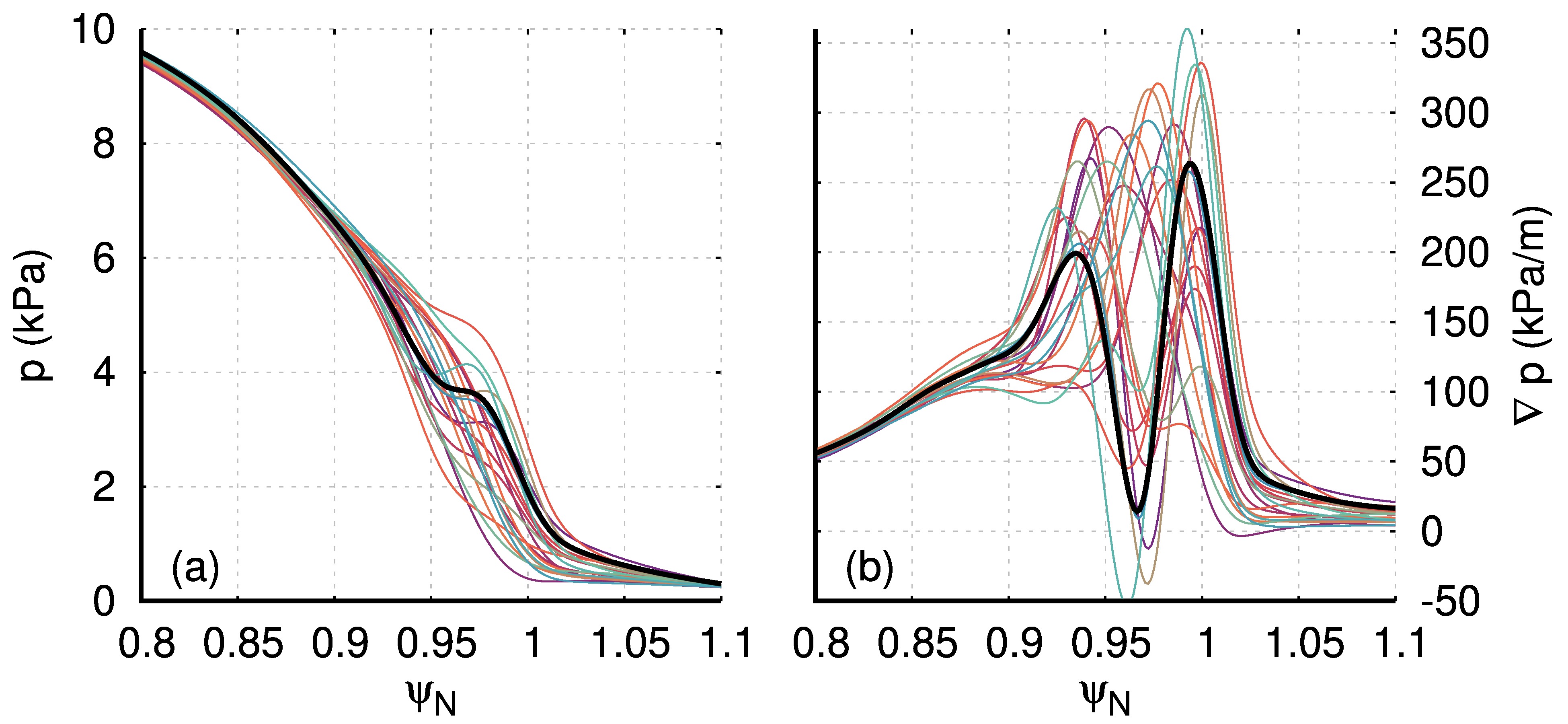}
    \caption{Pressure (a) and pressure gradient (b) profiles in the time window ${t=[4.0,6.0]~\mathrm{ms}}$ together with a time-averaged profile in black. The time-averaged profile shows a `staircase' structure with a large pressure gradient in the vicinity of the separatrix.}
    \label{fig:smallELM_p-dp_polline}
\end{figure}

Resistive peeling-ballooning modes that are destabilised below the ideal PB stability boundary regulate the pressure gradient about ${-250~\mathrm{kPa/m}}$. This is made clearer with pressure and pressure gradient profiles taken in the representative time frame of ${t=[4.0,6.0]~\mathrm{ms}}$ together with time-averaged profiles in black shown in fig.~\ref{fig:smallELM_p-dp_polline}(a) and (b). As mentioned before, the peeling-ballooning modes do not behave in a cyclical fashion, but cause a quasi-continuous power deposition in the inner and outer divertor targets, as shown in fig.~\ref{fig:smallELM_pressure_heat}(b).

\begin{figure}[!t]
    \centering
    \includegraphics[width=0.48\textwidth]{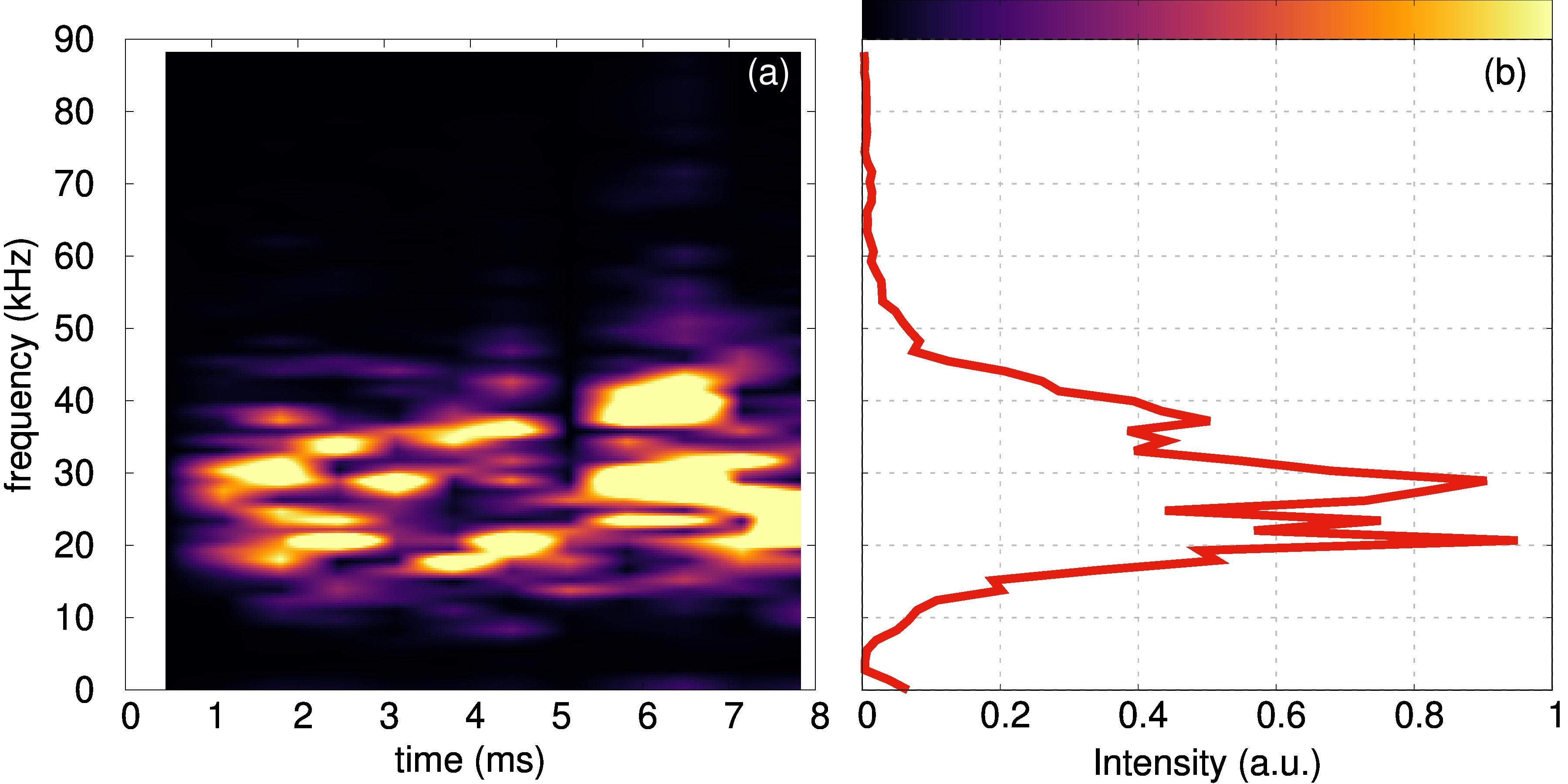}
    \caption{Time evolving (a) and averaged (b) frequency spectrogram of the temperature fluctuations in ${(R,Z)=(2.14,0.06)}$. Dominant frequencies in the range ${20-40~\mathrm{kHz}}$ can be observed in both cases.}
    \label{fig:smallELM_specgram}
\end{figure}

The fluctuating profile in the last $\sim7\%$ of the confined region is clearly visible in fig.~\ref{fig:smallELM_p-dp_polline}(a). It shows how the resistive PB modes regulate the pedestal in such a way that the steepness of the profiles cannot grow to large values. This is why these simulations feature only small ELMs and not a mixed regime with small ELMs and type-I ELMs. Taking the time-varying temperature fluctuations at a single point in the steep gradient region in the outer midplane, ${(R,Z)=(2.14,0.06)}$, a spectrogram is performed. As a result, a dominant frequency in the range of ${20-40~\mathrm{kHz}}$ is found, as can be observed in fig.~\ref{fig:smallELM_specgram}. A type-II ELMy H-mode in AUG with high triangularity and close to double null reported in Ref.~\cite{wolfrum2011characterization} was described as having an electron pressure gradient oscillating about ${\sim150~\mathrm{kPa/m}}$. The oscillating ${\nabla p_e}$ was reportedly caused by MHD modes which were associated with electromagnetic fluctuations observed in a wide radial extent peaking in a frequency range of ${30-50~\mathrm{kHz}}$. Both observations hint at qualitative similarities to the simulation results described in this section. Nevertheless, it must be noted that the present simulations were performed in a different magnetic configuration, \ie low triangularity and far from double null. Therefore, dedicated comparisons need to be performed in the future to produce quantitative comparisons between experiments and simulations. In particular, such comparisons will have to include variations of the plasma shape. 

\subsection{Divertor heat deposition}

\begin{figure}[!t]
    \centering
    \includegraphics[width=0.48\textwidth]{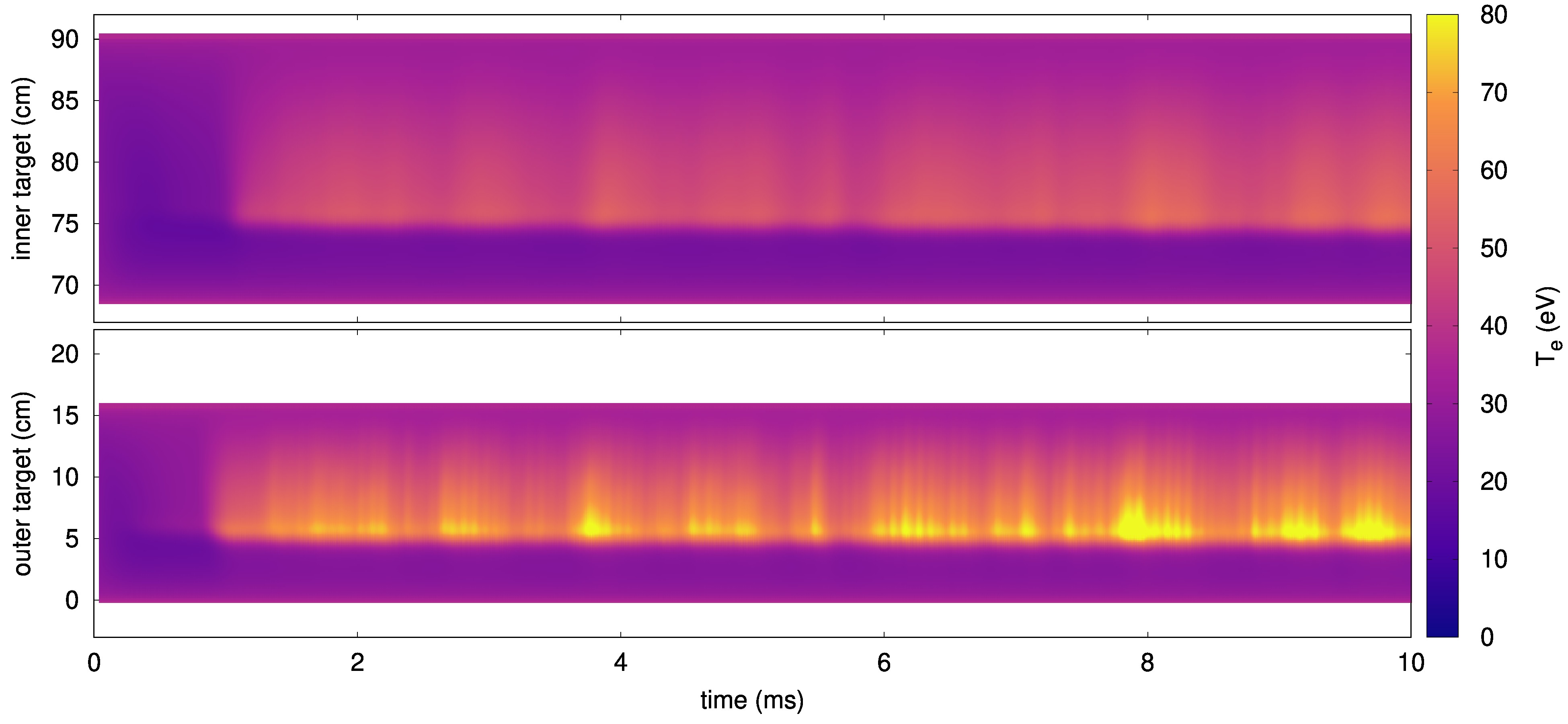}
    \caption{Time evolution of the inner (a) and outer (b) target electron temperature (${T_e=T/2}$) caused by the resistive PB modes excited at the very edge of the plasma. The inner target has a lower temperature than the outer target as well as a lower incident power.}
    \label{fig:smallELM_targets-Te}
\end{figure}

To show the quasi-continuous exhaust caused by the non-ideal peeling-ballooning modes excited near the separatrix, the electron temperature at the inner and outer divertor targets is plotted in fig.~\ref{fig:smallELM_targets-Te}(a) and (b), respectively. The target electron temperature is considered to be half of the plasma temperature and it is plotted for ${10~\mathrm{ms}}$ of simulation time. The inner divertor target has a lower target temperature than the outer divertor. Similarly, the incident power to the inner divertor is lower than to the outer divertor, as seen in fig.~\ref{fig:smallELM_pressure_heat}(b). There is a slight increase in the maximum target temperature (particularly visible in the outer target) as time progresses. This is due to the chosen heat source in the confined region which slowly increases the thermal energy content inside the separatrix. At any given time point, the heat deposition does not show significant variations in the toroidal direction. In other words, the heat deposition is roughly axisymmetric. It is important to note that the present simulations used only a simplified SOL transport model and, as such, the obtained heat distribution between targets will not necessarily reflect experimental observations. A more advanced SOL/divertor model is being developed in JOREK~\cite{Korving_2021eps} and it will be used for future simulations.

\section{Two simple paths to type-I ELMs}\label{sec:smallELMs_typeI}

Based on the simulations presented in the previous section, the heating power is increased to understand the response of the resistive PB modes. Increasing heating power causes the edge temperature (and its gradient) to increase, which, in turn, causes the local resistivity to decrease (${\eta\propto T^{-3/2}}$) and the pressure gradient and the diamagnetic drifts to grow larger. The stabilising influence of the diamagnetic effects and of $E_r$ (and its shear) onto PB modes becomes stronger and eventually completely stabilises them. At this point, the small ELM regime gives way to a type-I ELMy H-mode. Further details regarding this bifurcation determined by the applied heating power are presented in subsection~\ref{ssec:smallELMs-morePheat}. The transition from the small ELM regime to a type-I ELMy H-mode can also take place by sufficiently decreasing the separatrix density. Subsection~\ref{ssec:smallELMs-lowernsep} describes how decreasing $n_{e,\sep}$ (with respect to the pure small ELM simulations) manages to completely stabilise the resistive PB modes and gives way to a type-I ELMy H-mode. The decreasing separatrix density prompts three important stabilising effects to take place: a smaller edge pressure gradient, faster plasma flows since ${v^*_\mathrm{i}}$ and ${v_\mathrm{ExB}}$ are ${\propto1/n_i}$, and a higher bootstrap current density. 

\subsection{Increasing heating power}\label{ssec:smallELMs-morePheat}

\begin{figure}[!h]
    \centering
    \includegraphics[width=0.48\textwidth]{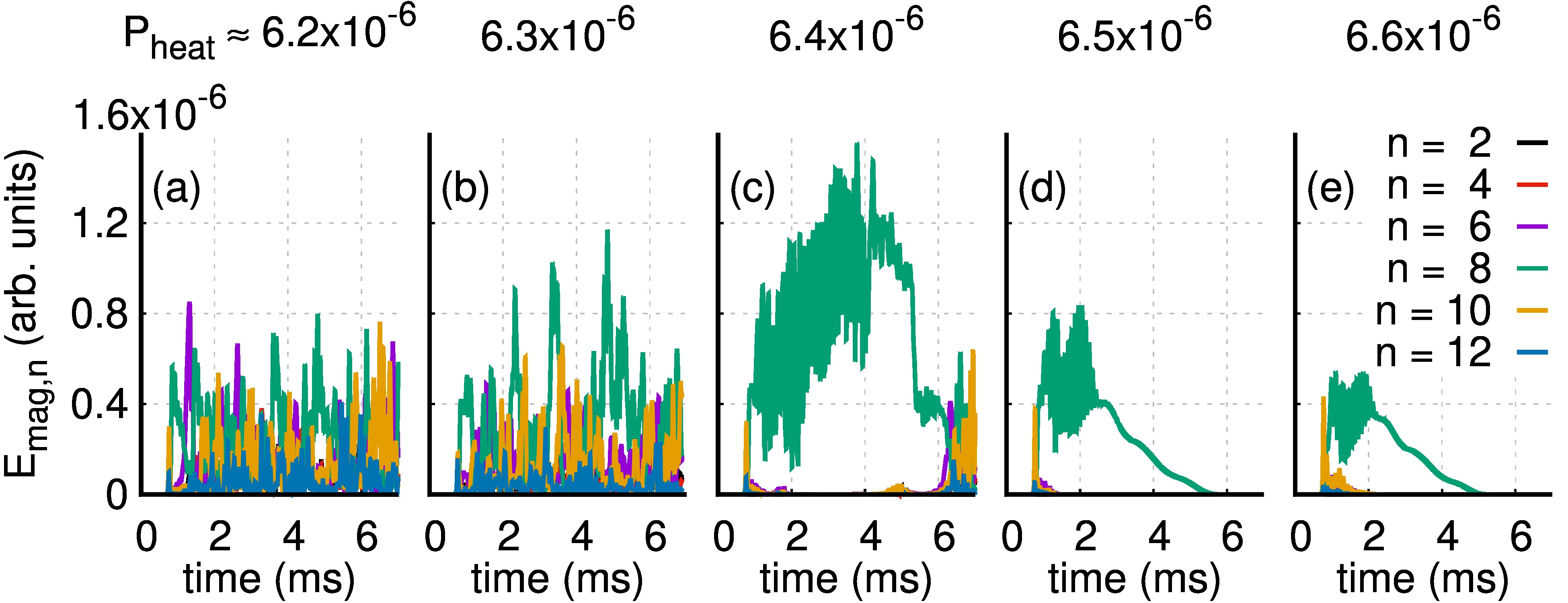}
    \caption{Magnetic energies of the non-axisymmetric perturbations for five different values of input heating power. The applied heating power increases progressively from (a)${(\approx13.0~\mathrm{MW})}$ to (e)${(\approx13.9~\mathrm{MW})}$. As a result of the increasing heating power, the magnetic energies of the $n>0$ perturbations become completely suppressed if sufficient additional heating power is considered.}
    \label{fig:smallELM_morePheat}
\end{figure}

The nominal heating power in JOREK units is ${6.2\times10^{-6}}$ (equivalent to ${\approx13~\mathrm{MW}}$) and it was applied in the simulation shown in fig.~\ref{fig:smallELM_linear_lownsep}. The magnetic energies of the non-axisymmetric perturbations for the first ${7~\mathrm{ms}}$ of said simulation are shown in fig.~\ref{fig:smallELM_morePheat}(a). In the subsequent sub-figures, the heating power is progressively increased\footnote{The excess heating power is always applied in the vicinity of the pedestal, such that the effect of the faster pedestal evolution can be rapidly determined. Depositing the excess heating power in the core produces the same results, but in a longer time scale as the excess heat needs to diffuse from the core to the pedestal top.} in small steps (the excess heating power is included from the beginning of each simulation). In fig.~\ref{fig:smallELM_morePheat}(b), the non-linear behaviour does not show many differences to the simulation with nominal $P_\heat$. However, in the next figure, a transient phase where the $n=8$ mode hosts most of the total non-axisymmetric energy, ${\Sigma_{n>0}} E_{\mathrm{mag}, n}$, is present. For this case with ${P_\mathrm{heat}\approx6.4\times10^{-6}}$ (${\approx13.5~\mathrm{MW}}$), the non-ideal PB modes become stabilised after roughly $10~\mathrm{ms}$. And in figs.~\ref{fig:smallELM_morePheat}(d) and (e), ${\Sigma_{n>0}} E_{\mathrm{mag}, n}$ is reduced until complete stabilisation. This mode stabilization allows the pedestal to build up and give rise to a type-I ELM crash eventually (discussed and shown in more detail later in this section).

\begin{figure}[!t]
    \centering
    \includegraphics[width=0.48\textwidth]{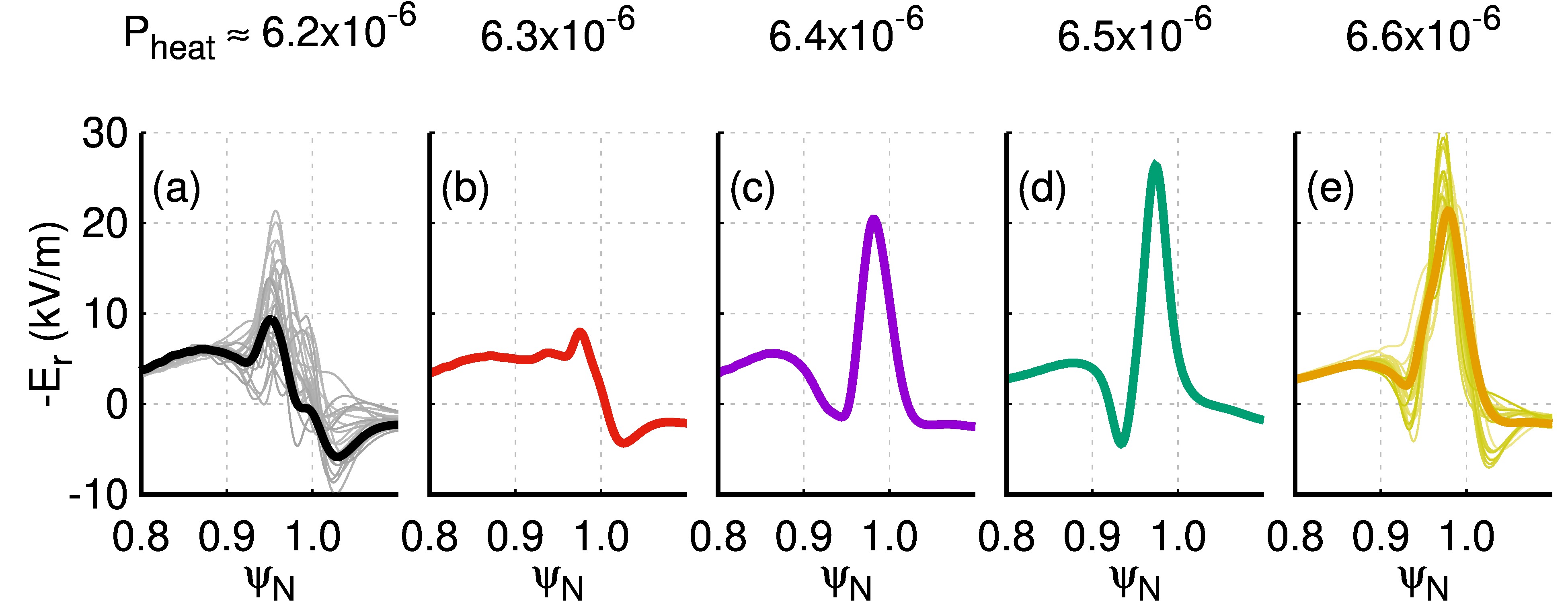}
    \caption{Time-averaged profiles of the outboard midplane radial electric field for five different values of input heating power. The applied heating power increases progressively from (a)${(\approx13.0~\mathrm{MW})}$ to (e)${(\approx13.9~\mathrm{MW})}$. }
    \label{fig:smallELM_morePheat_Er}
\end{figure}

To further understand what governs the transition from small ELMs to type-I ELMs, the radial electric field at the outboard midplane is averaged between ${1.0-2.0~\mathrm{ms}}$ and it is shown in figs.~\ref{fig:smallELM_morePheat_Er}(a)-(e). An interesting observation is that the three scenarios where the small ELMs become stabilised (c)-(e) have deeper radial electric field wells than the two cases that sustain the small ELMs (a) and (b). It is worth pointing out that even for the lowest heating power, the instantaneous radial electric field profiles at the outer midplane are often deeper than ${E_r\sim-15~\mathrm{kV/m}}$, a representative value which has been associated to the L-H transition in AUG~\cite{Sauter_2011,Cavedon_2020}. This can be seen in fig.~\ref{fig:smallELM_morePheat_Er}(a), which shows in gray the instantaneous profiles used to obtain the time-averaged profile (black). The scans with and without diamagnetic effects shown in section~\ref{ssec:smallELMs-extendedMHD-importance} together with the observations presented so far in this section indicate that small ELMs feel a stabilising effect from larger diamagnetic drifts and the resulting deeper radial electric field well.

\begin{figure}[!t]
    \centering
    \includegraphics[width=0.48\textwidth]{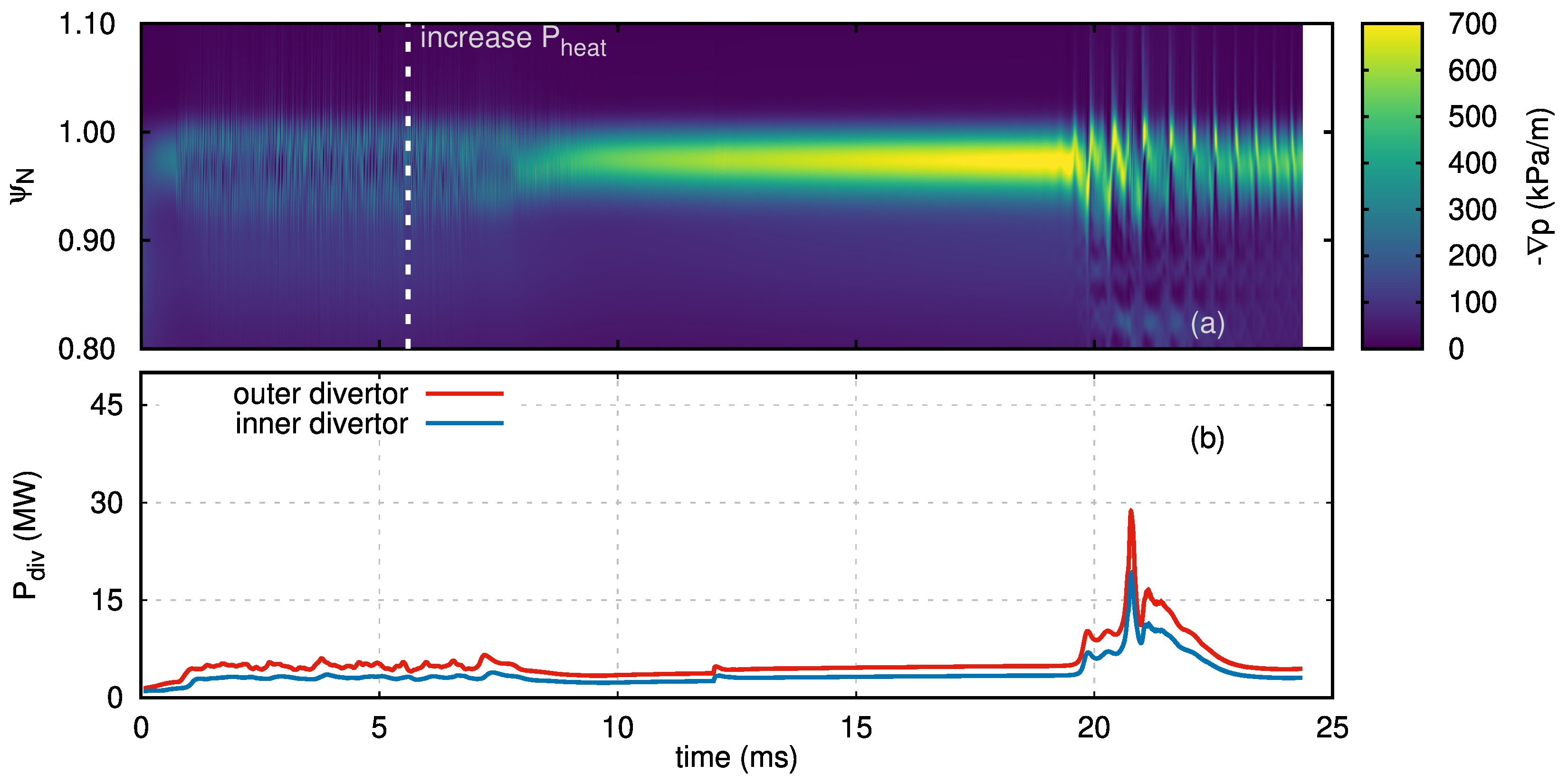}
    \caption{Time evolution of the outboard midplane pressure gradient at ${\varphi=0}$ (a) upon increasing the heating power at ${5.6~\mathrm{ms}}$. And the power incident on the inner and outer divertor targets (b) resulting from small ELMs (${t<10~\mathrm{ms}}$) and from a type-I ELM (${t\gtrsim19~\mathrm{ms}}$).}
    \label{fig:smallELM_increasePheat}
\end{figure}

\begin{figure}[!t]
    \centering
    \includegraphics[width=0.48\textwidth]{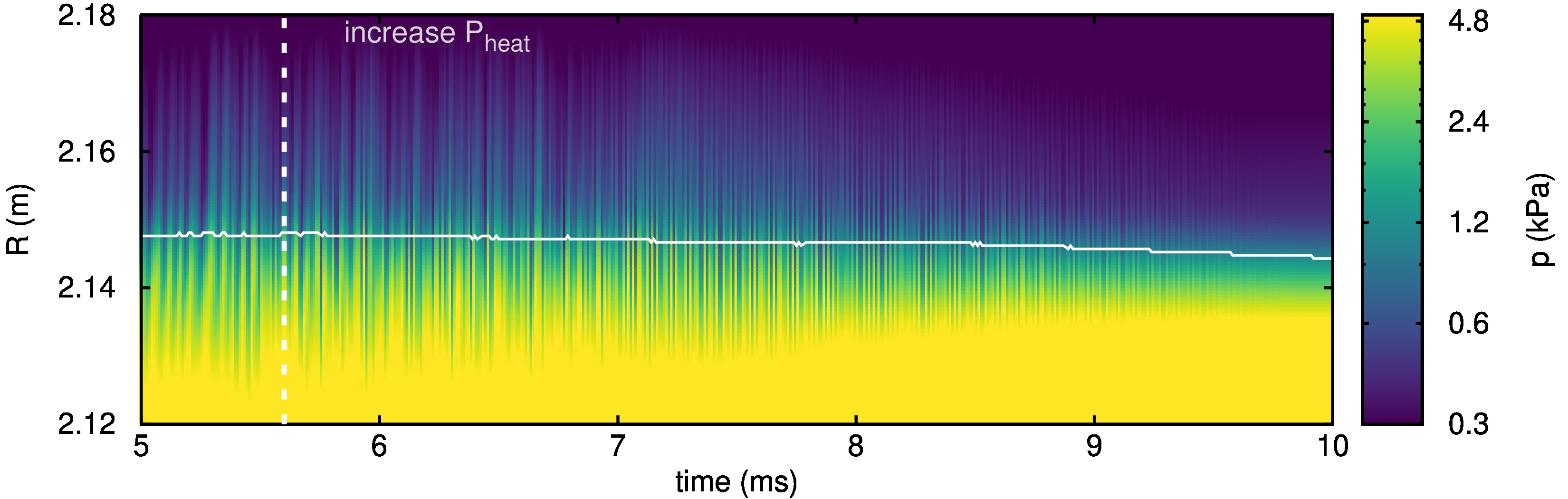}
    \caption{Time evolution of the outboard midplane pressure at ${\varphi=0}$ (a) upon increasing the heating power at ${5.6~\mathrm{ms}}$. The filaments expelled from the confined region become weaker when the heating power is increased. They ultimately completely disappear and the pedestal is able to grow further.}
    \label{fig:smallELM_increasePheat_zoom}
\end{figure}

In order to directly show the bifurcation from small ELMs to type-I ELMs, the heating power of the simulation described in section~\ref{ssec:smallELM_nonlinphase} is increased at ${5.6~\mathrm{ms}}$ of simulation time from ${\approx13.0}$ to ${\approx13.9~\mathrm{MW}}$. Resulting from the heating power increase, the resistive PB modes start to become smaller in amplitude and their radial extent starts to reduce. This process takes roughly ${4~\mathrm{ms}}$ to complete and, thereafter, a steeper pedestal is allowed to form. Ultimately, the steepening pedestal crosses the type-I ELM stability boundary and an ELM with dominant toroidal mode numbers ${n=2}$ and ${4}$ is excited. The process described in this paragraph can be evidenced in fig.~\ref{fig:smallELM_increasePheat}(a) and (b), which respectively show the ${\varphi=0}$ outboard midplane pressure gradient and the power incident on the inner and outer divertors. The small jump at ${t\approx12~\mathrm{ms}}$ takes place because the parallel heat conductivity had been increased roughly to the Spitzer-H\"arm values in this simulation, but the onset of the type-I ELM takes place regardless of said change. The incident power that reaches the divertor targets is significantly increased when the type-I ELM crash appears. Comparatively, it is clear that the small ELMs cause much weaker heat fluxes to the divertor targets. To directly show the influence of the increased heating power onto the resistive PB modes that cause small ELMs, the plasma pressure in real space together with the position of the separatrix are plotted for a restricted time frame between ${t=5}$ and ${10~\mathrm{ms}}$ in fig.~\ref{fig:smallELM_increasePheat_zoom}. The expelled filaments after the heating power increase seem to have smaller amplitudes and they travel for shorter distances. They eventually disappear completely. 


\begin{figure}[!t]
    \centering
    \includegraphics[width=0.48\textwidth]{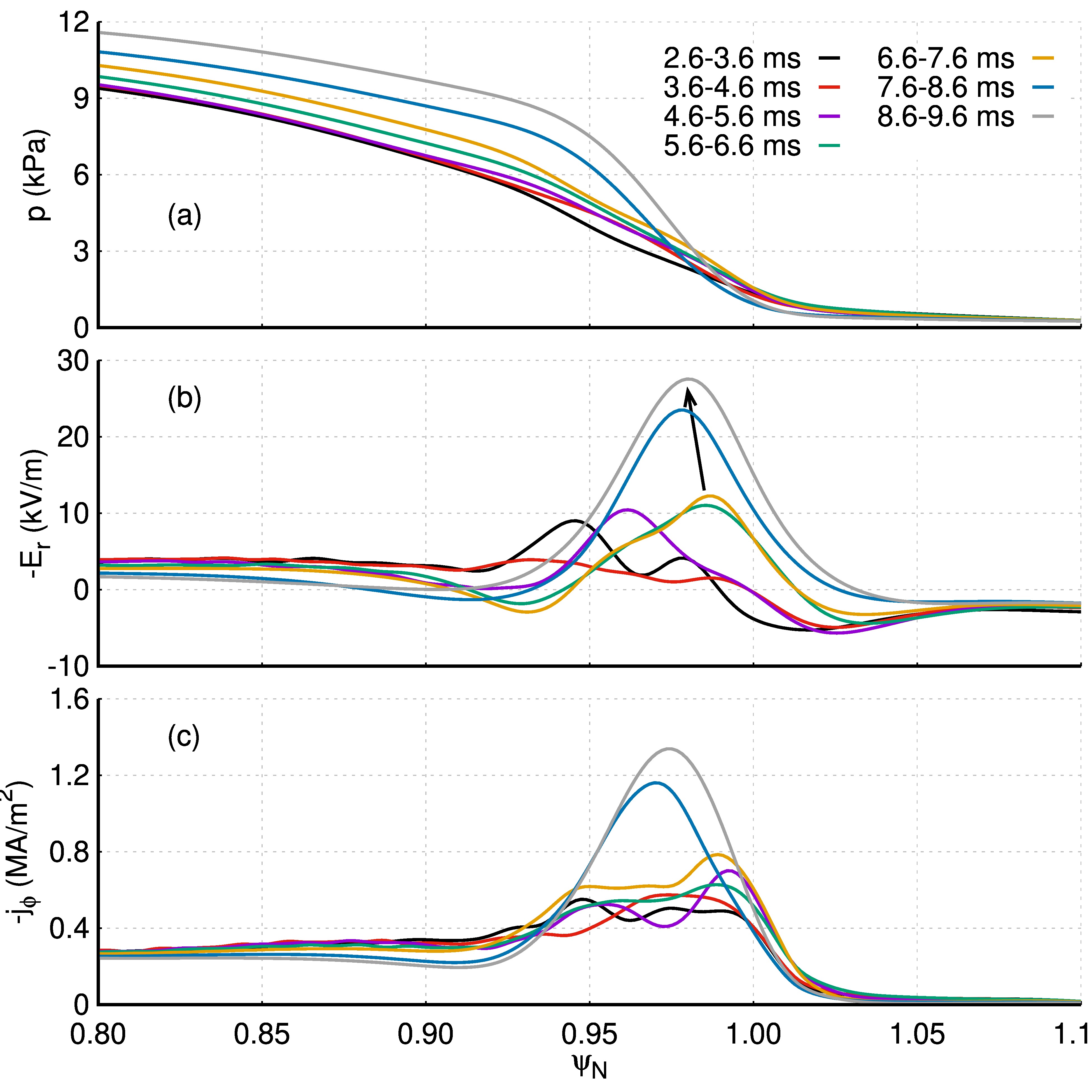}
    \caption{Evolution of the time-averaged ${\varphi=0}$ outboard midplane pressure (a), radial electric field (b), and toroidal current density (c). The time averaging is done for one millisecond intervals. The three time-averaged profiles before the heating power is increased have a shallow $E_r$, and the four profiles after ${P_\mathrm{heat}}$ show systematically higher ${p_\ped}$ and deeper $E_r$ well (as indicated by the black arrow for the latter).}
    \label{fig:smallELM_increasePheat_p-Er-jphi}
\end{figure}

In section~\ref{sec:AUGsmallELMs}, we discussed the response of AUG small ELMs at high separatrix density, low triangularity and high edge safety factor towards increasing heating power. Namely, in such experiments the small ELMs can become suppressed with sufficiently high additional heating power. The situation is similar for the EDA H-mode. For both cases, the role of plasma shaping in the suppression of small ELMs/QCM seems to be pivotal. In our simulations, which feature high ${n_{e,\sep}}$, low triangularity, and high ${q_{95}}$, the transition from a regime dominated by small ELMs towards a type-I ELM is obtained by suddenly increasing the input heating power in small ELM simulations. The small ELMs start to weaken and the filaments formed by the small ELMs are gradually reduced in amplitude until disappearing completely. ${p_\ped}$ rises with increasing ${P_\mathrm{heat}}$ due to the increase of ${T_\ped}$; ${n_{e,\ped}}$ remains unchanged in the first few milliseconds after the heating power was increased--- it only starts rising when the particle transport by small ELMs becomes significantly reduced. Additionally, due to the larger ${\nabla p}$, the radial electric field well at the plasma edge deepens, and the bootstrap current density starts to rise. Taking one millisecond time-averages, the outer midplane profiles are tracked during the pure small ELMs phase (from ${2.6~\mathrm{ms}}$ until ${5.6~\mathrm{ms}}$) and during the transition phase where the resistive PB modes start to disappear (from ${5.6~\mathrm{ms}}$ to ${9.6~\mathrm{ms}}$) and are plotted in fig.~\ref{fig:smallELM_increasePheat_p-Er-jphi}.

Three time-averaged profiles correspond to the original small ELM phase and show a roughly constant $p_\ped$, a weak $E_r$ at the edge, and low toroidal current density. On the other hand, the four profiles at higher heating power show systematically higher $p_\ped$, deeper $E_r$ well, and a broader and higher edge current density. The last time-averaged profiles in gray, with the highest $p_\ped$, ${\mathrm{min}(E_r)\approx-28~\mathrm{kV/m}}$ and a high toroidal current density, is taken during a phase that mostly has suppressed the resistive PB modes. The destabilising effect of the increased ${\nabla p}$ is tied to the stabilising effects of the deepening of the $E_r$ well and of the toroidal current density. The additional stabilising effect of the edge resistivity decreasing as the pedestal top temperature increases is also important at this stage. Therefore, the disappearance of the resistive PB modes appears to be due to the $E_r$ deepening, $-j_\varphi$ increasing, and the decreasing local $\eta$.

\subsection{Decreasing separatrix density}\label{ssec:smallELMs-lowernsep}

The previous section detailed the bifurcation from a small ELM-dominant regime to a type-I ELM by means of increasing the heating power. Another path towards type-I ELMs, starting from a small ELM regime, is to decrease the separatrix density. In the experiment, this can be achieved by reducing, or completely removing, the particle source given by a gas puff (replacing the particle flux by means of cryogenic deuterium pellet injection can keep $n_{e,\ped}$ approximately unchanged). Indeed, it has been reported that separatrix densities below ${\sim0.35 n_{GW}}$ do not host dominant small ELM phases~\cite{Labit_2019}. 

The simulations presented thus far had a separatrix density of ${n_{e,\sep}\approx3\times10^{19}~\mathrm{m^{-3}}}$. To reduce the separatrix density in our simulations, we reduce the edge density source; in order to maintain ${n_{e,\ped}}$ unchanged we increase the core density source. Resulting from the lower ${n_{e,\sep}}$, the pressure gradient is locally reduced, and it is overall shifted slightly inwards--- such response of the pressure profile position is also observed in experiments~\cite{Dunne_2016,Frassinetti_2019}. The decrease of separatrix density (at fixed ${n_{e,\ped}}$) also causes a deeper ${E_r}$ well (because ${E_r\propto1/n_e}$) and a higher bootstrap current density (because the density gradient increases). An important caveat must be mentioned: many physical effects related to the pedestal position (particularly related to neutrals penetration) are not included in the JOREK model used for these simulations and, therefore, the qualitative agreement will likely not translate to a quantitative agreement at this stage.

\begin{figure}[!t]
    \centering
    \includegraphics[width=0.48\textwidth]{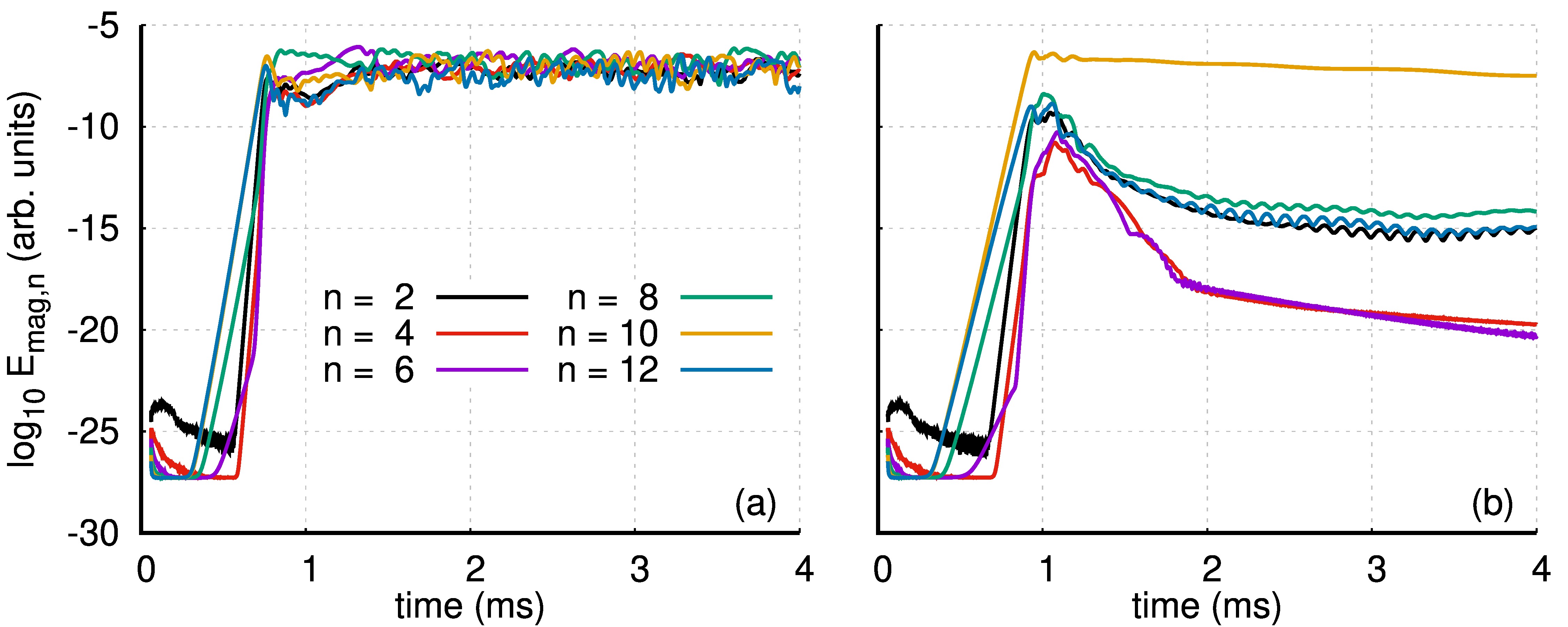}
    \caption{Magnetic energies of the non-axisymmetric modes in logarithmic scale in the first $4$ milliseconds of simulation time for (a) small ELMs at high ${n_\sep(\sim3\times10^{19}~\mathrm{m^{-3}})}$ and (b) their response to lower separatrix density (${\sim2\times10^{19}~\mathrm{m^{-3}}}$).}
    \label{fig:smallELM_linear_lownsep}
\end{figure}

A new simulation is then set-up with lower ${n_{e,\sep}=2\times10^{19}~\mathrm{m^{-3}}}$. The non-zero toroidal modes are included after ${0.1~\mathrm{ms}}$ (exactly the same time as the small ELM simulations described in section~\ref{sec:smallELM_n!0}). The magnetic energies of the high and low ${n_{e,\sep}}$ simulations are shown in fig.~\ref{fig:smallELM_linear_lownsep}(a) and (b), respectively. The linear phases are similar between the two cases, with growing $n=8,10,12$ high-$n$ (peeling-)ballooning modes. But the simulation with low $n_{e,\sep}$ (b) deviates as the ${n<10}$ and ${n=12}$ become completely stabilised, and the ${n=10}$ only reaches small amplitudes and does not affect the $n=0$ background. The stabilisation of the modes with higher toroidal mode numbers leads to a single toroidal mode number with very small amplitude that does not cause any changes to the background plasma.

\begin{figure}[!t]
    \centering
    \includegraphics[width=0.48\textwidth]{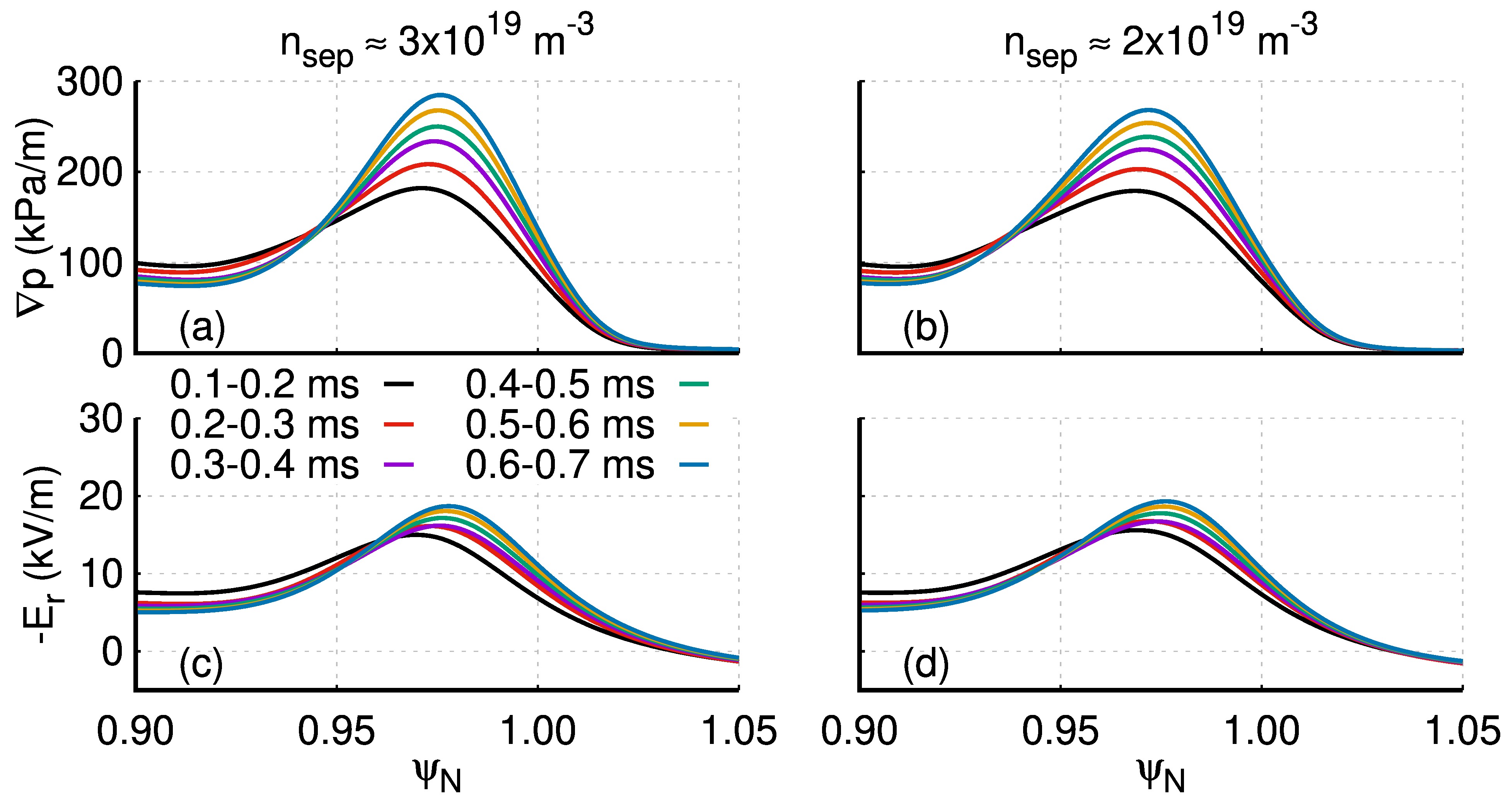}
    \caption{Time-averaged outboard midplane profiles of the pressure gradient and $E_r$ during the linear growth phase for nominal separatrix density (a) and (c), and for lower separatrix density (b) and (d).}
    \label{fig:smallELM_dp-Er_lownsep}
\end{figure}

The simulation at lower separatrix density, fig~\ref{fig:smallELM_linear_lownsep}(b), sees the pedestal evolve, but it was not continued until a type-I ELM is reached to save computing time. Time-averaged outer midplane profiles of the pressure gradient and radial electric field are displayed in fig.~\ref{fig:smallELM_dp-Er_lownsep} for the small ELMs with ${n_{e,\sep}\approx3\times10^{19}~\mathrm{m^{-3}}}$ (a) and (c), and for the lowered separatrix density case (${n_{e,\sep}\approx2\times10^{19}~\mathrm{m^{-3}}}$) (b) and (d). The profiles are averaged over $0.1~\mathrm{ms}$ during the linear growth phase, $<0.7~\mathrm{ms}$. The lower $n_\sep$ causes an inward shift of $\nabla p$ and, particularly, a smaller pressure gradient in the vicinity of the separatrix. It additionally allows for a deeper $E_r$ well and an increase of the bootstrap current density (not shown). Diminishing the destabilising influence of the large $\nabla p$ near the separatrix together with the stabilising influence of the deeper $E_r$ and the higher $-j_\varphi$ cause the small ELMs to become completely stabilised.

\section{Conclusions}\label{sec:smallELM_conclusions}

H-mode operation without large type-I ELMs is an imperative requirement for ITER in high-performance conditions. To this purpose, naturally ELM-free H-modes and ELM mitigated/suppressed regimes are considered and actively researched. In AUG, several such alternatives are under investigation; two of them are the quasi-continuous exhaust (QCE) regime and the enhanced D-alpha H-mode. Both can be operated completely without type-I ELMs. The pedestal is limited by small ELMs for the QCE regime and by a quasi-coherent mode for the EDA H-mode. These transport mechanisms quasi-continuously expel heat and particles from the confined region. It is presently unclear whether or not it will be possible to operate such regimes in ITER. Simulations performed with JOREK, which show several key features of small ELMs have been presented in this paper. Resistive peeling-ballooning modes near the separatrix are identified as the transport mechanism underlying such small ELMs. 

Modelling the pedestal build-up at fixed pedestal width, with stationary diffusion coefficients and sources, resistive peeling-ballooning modes that regulate the pedestal below the ideal PB stability boundary are observed under appropriate conditions. Namely, simultaneously high separatrix density and not too much heating power. The necessary conditions to sustain sufficient outwards transport by small ELMs is primarily determined by the separatrix density and the input heating power. In particular, simulations with high $n_{e,\sep}$ and low heating power observe phases (longer than ${10~\mathrm{ms}}$) with quasi-continuous outwards transport that prevent the pedestal from reaching a type-I ELM unstable scenario. The resistive nature of such PB modes is determined by the fact that they appear below the ideal PB stability boundary, because their growth rates are largely reduced/enhanced by decreasing/increasing resistivity, and because their poloidal mode velocities are measured to be close to that expected for resistive modes. 

An important ingredient required in order to properly simulate these resistive PB modes is the inclusion of diamagnetic effects, which (in the simulations) allow the radial electric field well to develop in the pedestal region. In the absence of diamagnetic effects, it is not possible to stabilise the small ELMs by increasing the heating power. In contrast, when diamagnetic effects are included, the small ELMs become completely stabilised and the plasma state moves to a type-I ELMy H-mode by increasing ${P_\mathrm{heat}}$. Similarly, decreasing the separatrix density completely stabilises the small ELMs if diamagnetic effects are included. Another important effect that should be included when modelling these instabilities is the bootstrap current density because it has a stabilising influence onto high-$n$ peeling-ballooning modes. At the moment, JOREK evolves the bootstrap current density through the Sauter formula~\cite{Sauter1999,Sauter2002}, as explained in section~\ref{sec:smallELM_setup}. However, the Sauter expression is known to be inaccurate depending on the parameter regime, particularly at high collisionality~\cite{Belli_2014}. Therefore, an improvement of the bootstrap current density source in JOREK will be pursued in the future. Finally, the simplified resistivity with only Spitzer temperature dependency used in JOREK should be improved to include the influence of neoclassical effects and effective main ion charge greater than unity.

\section*{Acknowledgements}

This work has been carried out within the framework of the EUROfusion Consortium and has received funding from the Euratom research and training program 2014-2018 and 2019-2020 under grant agreement No 633053. The views and opinions expressed herein do not necessarily reflect those of the European Commission. In particular, contributions by EUROfusion work packages Enabling Research (EnR) and Medium Size Tokamaks (MST) is acknowledged. The simulations presented in this work were performed on the Marconi-Fusion supercomputer operated by CINECA in Italy.

\bibliography{main}

\end{document}